\documentclass[12pt,tightenlines,nofootinbib]{revtex4-2}
\usepackage{amsfonts,amsmath,amssymb}
\usepackage{enumitem}
\usepackage{txfonts}
\usepackage{graphicx,subfigure,wrapfig}
\usepackage{color}
\usepackage{lipsum,booktabs}
\usepackage{floatrow,float}
\usepackage{epstopdf}
\usepackage{hyperref}
\usepackage[normalem]{ulem}
\usepackage{wasysym}

\DeclareMathOperator{\swr}{swr}

\begin{document}

\title{Pattern dynamics and stochasticity of the brain rhythms}
\author{C. Hoffman$^1$, J. Cheng$^2$, D. Ji$^2$, Y. Dabaghian$^1$}
\affiliation{$^1$Department of Neurology, The University of Texas McGovern Medical School, 6431 Fannin St, Houston, TX 77030\\
$^2$Department of Neuroscience, Baylor College of Medicine, Houston, TX 77030,\\
$^{*}$e-mail: Yuri.A.Dabaghian@uth.tmc.edu}
\date{\today}

\begin{abstract}
	Our current understanding of brain rhythms is based on quantifying their instantaneous or
	time-averaged characteristics. What remains unexplored, is the actual structure of the
	waves---their shapes and patterns over finite timescales. To address this, we used two 
	independent approaches to link wave forms to their physiological functions: the first is
	based on quantifying their consistency with the underlying mean behavior, and the second
	assesses ``orderliness" of the waves' features. The corresponding measures capture the wave's
	characteristic and abnormal behaviors, such as atypical periodicity or excessive clustering, 
	and demonstrate coupling between the patterns' dynamics and the animal's location, speed and
	acceleration.
	Specifically, we studied patterns of $\theta$ and $\gamma$ waves, and Sharp Wave Ripples, and
	observed speed-modulated changes of the wave's cadence, an antiphase relationship between 
	orderliness and acceleration, as well as spatial selectiveness of patterns. Furthermore, we
	found an interdependence between orderliness and regularity: larger deviations from steady 
	oscillatory behavior tend to accompany disarrayed temporal cluttering of peaks and troughs. 
	Taken together, our results offer a complementary---mesoscale---perspective	on brain wave 
	structure, dynamics, and functionality.
\end{abstract}
\maketitle
\newpage

\section{Introduction}
\label{sec:intro}

Common approaches to studying ``brain rhythms" can be broadly divided in two categories. The first is based
on correlating the wave's instantaneous phases, amplitudes and frequencies with parameters of cognitive, 
behavioral or neuronal processes. For example, instantaneous phases of $\theta$-wave were found to modulate
neuronal spikings \cite{Skaggs,Benchenane}, while $\gamma$-waves' amplitudes and frequencies link the synaptic
\cite{Nikoli,ClgMsr,ColginGm} and circuit \cite{LisBuz,Lismn1,Lismn3} current flow to learning dynamics. 
The second category of analyses is based on quantifying the brain waves' time-averaged characteristics, e.g.,
establishing dependencies between the mean $\theta$-frequency and the animal's speed \cite{Richard,BuzTheta2}
or acceleration \cite{Kropff} or linking rising mean $\theta$- and $\gamma$-power to heightened attention 
states \cite{Rangel,Kropff} and so forth.

\begin{figure}[h]
	\centering
	\includegraphics[scale=0.84]{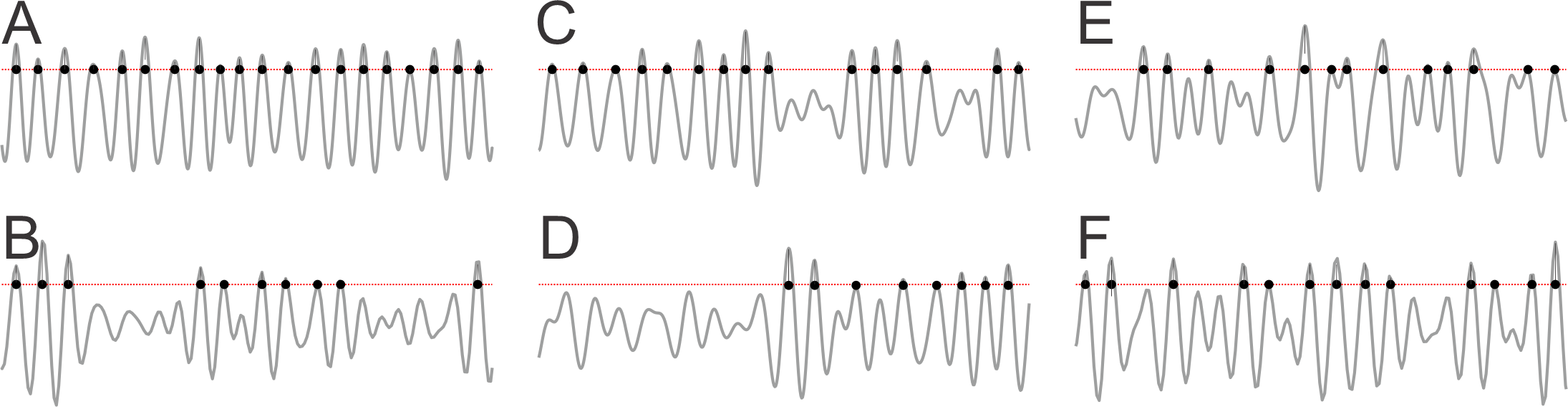}
	\caption{{\footnotesize
			\textbf{Waveform morphologies}. \textbf{A}. A wave exhibiting a nearly periodic sequence of peak is
			commonly found among $\theta$-oscillations, but rarely in higher-frequency waves. Conversely, the
			intermittent patterns shown on panel \textbf{B} may be exhibited by fast $\gamma$ or sharp waves,
			but is atypical for lower-frequencies. The temporal clustering shown on panel \textbf{C} is not 
			unusual for the $\gamma$-wave, but the uneven pattern on panel \textbf{D} would be an irregularity.
			In contrast, the patterns shown on panels \textbf{E} and \textbf{F} are all in all ordinary and
			seem to fluctuate with a fixed oscillatory rate. Shown are the peaks exceeding one standard deviation
			from the mean.
	}}
	\label{fig:pat}
\end{figure}

However, little work has been done to examine the waves' overall shape, e.g., the temporal arrangement of 
peaks and troughs, or sequences of ripples or spindles generated over finite periods. Yet, the physiological
relevance of the brain wave morphology is well recognized: rigidly periodic or excessively irregular rhythms
that contravene a certain ``natural" level of statistical variability are suggestive of circuit pathologies
\cite{Wilkinson,Donoghue,Eissa,Nicola,Hawkins,Blanco,Cole} or may indicate external driving \cite{Mostafa,Will}.
For example, the nearly periodic sequence of peaks shown on Fig.~\ref{fig:pat}A is common for $\theta$-waves,
but certainly too orderly for the $\gamma$-waves. Conversely, the intermittent pattern on Fig.~\ref{fig:pat}B
is unlikely to appear among $\theta$-waves, but may represent irregular $\gamma$-activity or typical sharp waves.
On the other hand, the series of clumping peaks shown on Fig.~\ref{fig:pat}C seem usual for in $\gamma$ waves,
but the cluttered pattern on Fig.~\ref{fig:pat}D may be a manifestation of a particular process. In contrast,
the waveforms shown on Fig.~\ref{fig:pat}E,F appear to represent a mundane level of arhythmicity and temporal 
disorder expected in most $\theta$ and $\gamma$ waves.

It remains unclear however, how to identify these patterns impartially, how to quantify the intuitive notions 
of ``regularity," ``typicality," ``orderliness," etc. Furthermore, since all brain waves exhibit a certain 
level of erraticness, it is unclear how justified are the experiential, visceral classifications of the 
waveforms. For example, might the ``improbable" patterns illustrated above be attributed to mere fluctuations
of otherwise regular waves, or should they be viewed as a structural peculiarity? 

In the following, we address these questions in the context of two independent mathematical frameworks, using
two cognate pattern quantifications that allow understanding the brain rhythms' functional structure at
intermediate timescales and their role in behavior and cognition.

\section{Approach}
\label{sec:app}
\begin{figure}[h]
	\centering
	\includegraphics[scale=0.84]{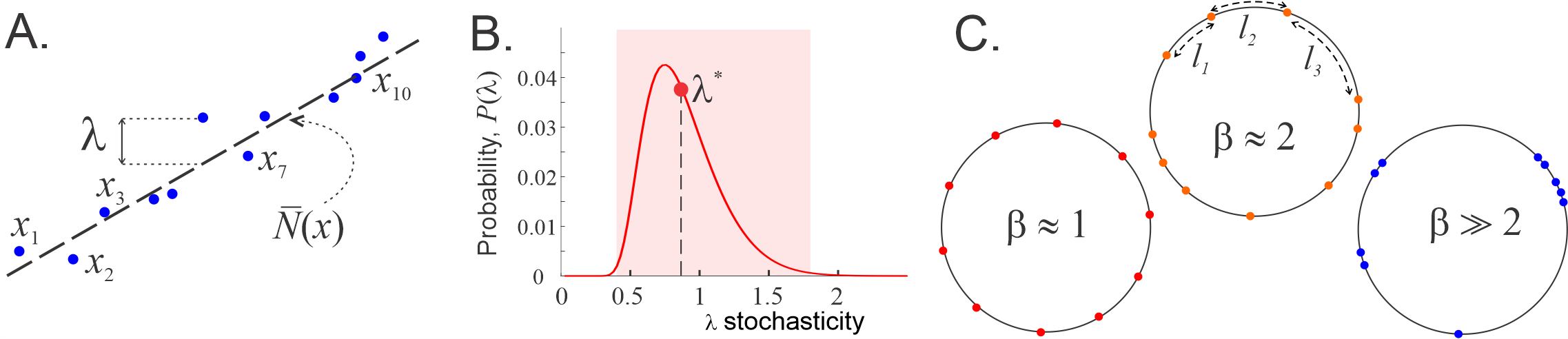}
	\caption{{\footnotesize
			\textbf{Stochasticity parameters}. \textbf{A}. The elements of an ordered sequence $X=\{x_1,x_2,
			\ldots,x_n\}$ following a linear trend $\bar{N}(x)=mx+b$ (dashed line). The sequence's deviations
			from the mean, $\lambda(X)$, exhibit statistical universality and can hence impartially 
			characterize the stochasticity of the individual data sequences $X$.
			\textbf{B}. The probability distribution of $\lambda$-scores is unimodal, with mean $\lambda^{\ast}
			\approx 0.87$ (red dot).	About $99.7\%$ of sequences produce $\lambda$-scores in	the interval
			$0.4\leq\lambda(X)\leq 1.8$ (pink stripe); these sequences are typical and consistent with the 
			underlying mean behavior. In contrast, sequences with smaller or larger $\lambda$-scores are 
			statistically uncommon. 
			\textbf{C}. A sequence $X$ arranged on a circle of length $L$ produces a set of $n$ arcs. The 
			normalized quadratic sum of the arc lengths is small for orderly sequences, $\beta\approx1$ (left),
			as high as $\beta\approx n$ for the ``clustering" sequences (right), and intermediate, $\beta\approx
			2$ (middle), for generic sequences. 
	}}
	\label{fig:stochs}
\end{figure}

\textit{1. Kolmogorov stochasticity, $\lambda$}, describes deviation of an ordered sequence, $X$, from the 
overall trend, and a remarkable observation made in \cite{Kolmogorov} is that this score is universally 
distributed. As it turns out, deviations $\lambda(X)$ that are too high or too low are rare: sequences with
$\lambda(X)\leq0.4$ or $\lambda(X)\geq 1.8$ appear with probability less than $0.3\%$ (Fig.~\ref{fig:stochs}A,B, 
\cite{Stephens,Arnold1,Arnold2,Arnold3,Arnold4,Arnold5}).
In other words, typical patterns are consistent with the underlying mean behavior and produce a limited range
of $\lambda$-values, with mean $\lambda^{\ast}\approx 0.87$. Thus, the $\lambda$-score can serve as a universal
measure of \textit{stochasticity}\footnote{Throughout the text, terminological definitions and highlights are 
	given in \textit{italics}.} and be used for identifying statistical biases (or lack of thereof) in various 
patterns \cite{KolMen,Stark,Gurzadyan1,Gurzadyan2,Brandouy,Ford}.

\textit{2. Arnold stochasticity, $\beta$}, is alternative measure that quantifies whether the elements of a
pattern ``repel''  or ``attract" each other. Repelling elements seek to maximize separations and hence produce
orderly, more equispaced arrangements, while attracting elements tend to cluster together. As shown
in \cite{ArnoldB1,ArnoldB2,ArnoldB3,ArnoldB4}, for ordered patterns $\beta(X)\approx1$, for clustering ones
$\beta(X)$ can be high, while sequences with independent elements yield $\beta$-values close to the impartial 
mean, $\beta^{\ast}\approx2$ (Fig.~\ref{fig:stochs}C, Suppl. Sec.~\ref{sec:met}). Thus, the $\beta$-score can
be used to characterize \textit{orderliness} of brain rhythms \cite{ArnoldB1,ArnoldB2,ArnoldB3,ArnoldB4}, 
complementing the $\lambda$-score. 

\textit{3. Time-dependence}. The recurrent nature of brain rhythms suggests dynamic generalization of $\lambda$ 
and $\beta$. Given a time window, $L$, containing a sequence of events, $X_t$, such as $\theta$-peaks or Sharp 
Wave Ripples (SWR), evaluate the parameters $\lambda(X_t)$ and $\beta(X_t)$, then shift the window over a time
step $\Delta t$, evaluate the next $\lambda(X_{t+\Delta t})$ and $\beta(X_{t+\Delta t})$, and so on. The 
consecutive segments, obtained by small window shifts, $X_t, X_{t+\Delta t}, X_{t+2\Delta t},\ldots$, differ 
only slightly from one another. The resulting time-dependencies $\lambda(t)$ and $\beta(t)$ will define the
dynamics of stochasticity over the signal's entire span.

For a visualization, one can imagine the elements of a given sample sequence, $X_{t+k\Delta t}$, as ``beads" 
scattered over a necklace of length $L$. As the sliding window shifts forward in time, the beads shift back and
may disappear at the back of the window, and new beads may appear toward the front, while a majority of the beads
retain their relative positions. The corresponding $\lambda$- and $\beta$-values will then produce semi-continuous
time-dependencies $\lambda(t)$ and $\beta(t)$ that quantify the ``necklace dynamics"---gradual pattern changes.
The parameter $\beta$ then describes the orderliness of the beads' distribution over the necklace, while
$\lambda$ measures how typical the beads' arrangement is overall. 

\begin{figure}[h]
	\centering
	\includegraphics[scale=0.8]{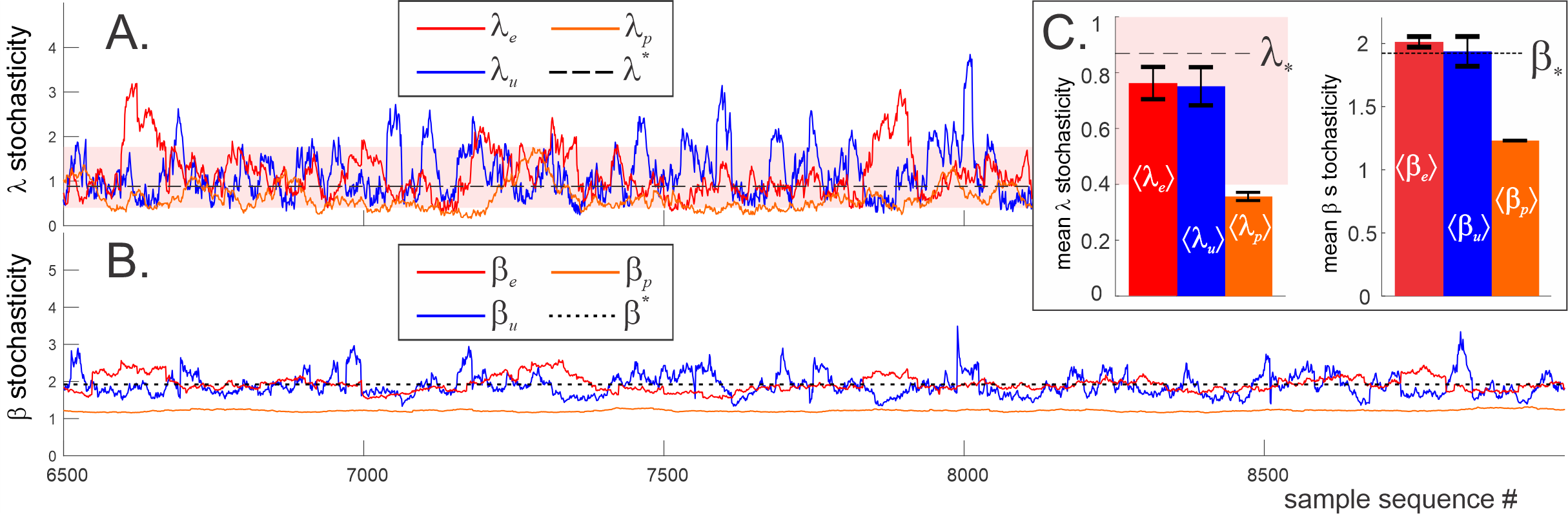}
	\caption{{\footnotesize
			\textbf{Pattern dynamics for three kinds of random sequences} in which the intervals between
			consecutive points are distributed 1) exponentially with the rate $\nu=2$; 2) uniformly
			with constant density $\rho=1$; or 3) with Poisson rate $\mu=5$. Sample intervals are selected
			proportionally to the distribution scales ($L_u=25\rho$, $L_e=25\nu$, and $L_p=25\mu$, so that each
			sample sequence contains about $n=25$ elements) and are shifted by a single data point at a time. 
			\textbf{A}. The Kolmogorov parameter of the exponential sequence (red trace, $\lambda_e$), uniform
			sequence (blue trace, $\lambda_u$) and Poisson sequence (orange trace, $\lambda_p$) remain mostly
			within the ``pink zone" of stochastic typicality (same pink stripe as on Fig.~\ref{fig:stochs}B). 
			$\lambda_u$ is the most volatile and often escapes the expected range, whereas $\lambda_p$ is more
			compliant, lingering below the expected mean $\lambda_p\lesssim\lambda^{\ast}\approx 0.87$ (black
			dashed line).		
			\textbf{B}. The corresponding Arnold stochasticity parameters show similar behavior: $\beta_u=1.93
			\pm 0.2$ fluctuates around the expected mean $\beta^{\ast}(25) = 1.92$ (black dotted line). The 
			exponential sequence has smaller $\beta$-variations and a slightly higher mean, $\beta_e=2\pm 0.04$.
			The Poisson sequence is the least stochastic (nearly-periodic), with $\beta_p=1.22\pm 0.004$, due 
			to statistical suppression of small and large gaps.
			\textbf{C}. The mean stochasticity scores computed for about $10^4$ random patterns of each type.}
	}
	\label{fig:rndstoch}
\end{figure}

As an illustration, consider a data series $X$ with random spacings between the adjacent values---intervals
drawn from exponential, uniform, and Poisson distributions, with sample subsequences containing about $25$
consecutive elements. As shown on Fig.~\ref{fig:rndstoch}A,C, the $\lambda(t)$-dependence of the exponential
sequences remains, for the most part, constrained within the ``typicality band" (pink stripe on 
Figs.~\ref{fig:stochs}B and Fig.~\ref{fig:rndstoch}A), while the uniformly distributed patterns are more 
variable and Poisson patterns follow the mean most closely. The $\beta$-scores of exponentially and uniformly
distributed patterns are overall mundane, while the Poisson patterns exhibit periodic-like orderliness. 

The mean $\lambda$- and $\beta$-scores in the uniform and the exponential sequences are close to universal
means, $\lambda^{\ast}$ and $\beta^{\ast}$, which shows that, on average, they are statistically unbiased.
In contrast, the Poisson-distributed patterns are atypically orderly, due to statistically suppressed small
and large gaps between neighboring elements (Fig.~\ref{fig:rndstoch}B).

The fluctuations of stochasticity scores---the rises and drops of $\lambda(t)$ and $\beta(t)$ dependencies on
Fig.~\ref{fig:rndstoch}---are chancy, since random sequences vary sporadically between instantiations. In 
contrast, brain wave patterns may carry physiological information, and the dynamics of their stochasticity
may serve as an independent characterization of circuit activity at a mesoscopic timescale in different 
behavioral and cognitive states, as discussed below.

\section{Results}
\label{sec:res}

\subsection{Stochasticity in time}
We analyzed Local Field Potentials (LFP) recorded in the hippocampal CA1 area of wild type male 
mice\footnote{The data used in this work was outlined in \cite{ChengJi}.} and studied their $\theta$-wave,
$\gamma$-wave, and SWR patterns \cite{ColginR}. The recurring nature of brain rhythms suggests that their key
features distribute uniformly over sufficiently long periods. Therefore, the expected mean used for evaluating
the Kolmogorov $\lambda$-parameter is linear,
\begin{equation}
\bar{N}(t) = m t + b,
\label{lin}
\end{equation}
with the coefficients $m$ and $b$ obtained via linear regression. The lengths of the sample sequences 
were then selected to highlight the specific wave's structure and functions, as described below.

\begin{figure}[h]
	\centering
	\includegraphics[scale=0.8]{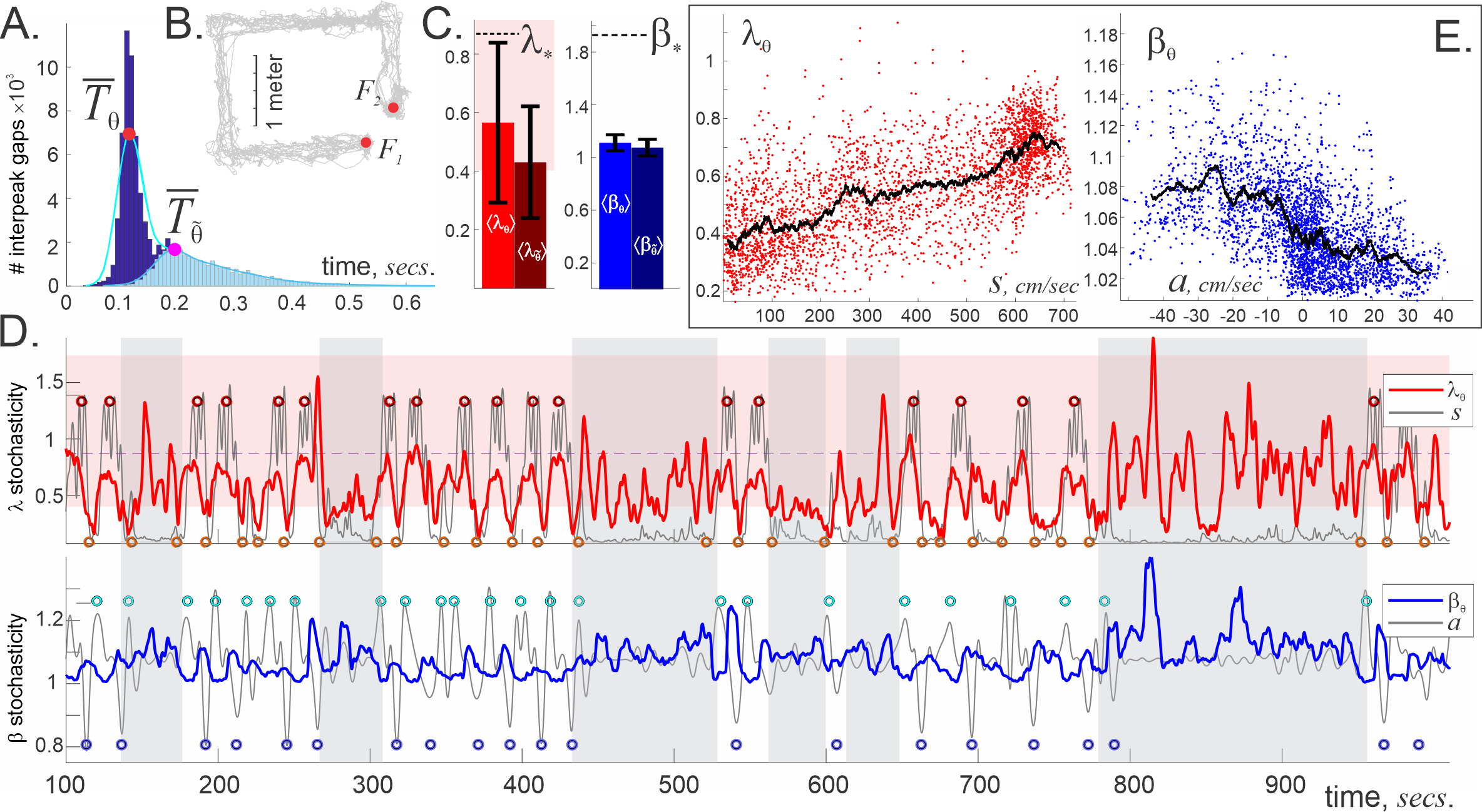}
	\caption{{\footnotesize
			\textbf{$\theta$-wave's stochasticity}. 
			\textbf{A}. A histogram of intervals between subsequent $\theta$-peaks concentrates around the
			characteristic $\theta$-period, $\overline{T}_{\theta}\approx 110$ msec: gaps shorter than 
			$\overline{T}_{\theta}/2$ or wider than $2\overline{T}_{\theta}$ are rare. $\theta$-amplitude, 
			$\tilde{\theta}$, oscillates with $T_{\tilde{\theta}}\approx 180$ msec period.
			\textbf{B}. The animal's lapses (trajectory shown by gray line) between food wells, $F_1$ and 
			$F_2$, take on average $22$ secs.
			\textbf{C}. Due to quasiperiodicity of the $\theta$-wave and of its envelope, $\tilde{\theta}$, 
			the average scores $\langle\lambda_{\theta}\rangle$, $\langle\tilde{\lambda}_{\theta}\rangle$, 
			$\langle\beta_{\theta}\rangle$ and $\langle\tilde{\beta}_{\theta}\rangle$ are significantly 
			lower than the impartial means $\lambda^{\ast}$ and $\beta^{\ast}$, with small deviations (data 
			for $5$ mice). 
			\textbf{D}. The dynamics of $\lambda_{\theta}(t)$ (red trace, upper panel) correlates with the 
			speed profile (gray line) when the mouse moves methodically. The $\lambda_{\theta}(t)$-stochasticity
			remains mostly within the ``typical" range (pink stripe in the background), falling below it as the
			mouse slows down. For rapid moves there is a clear similarity between the $\lambda_{\theta}$-score
			and the speed, e.g., their peaks and troughs roughly match. When the mouse meanders (vertical gray
			stripes), the coupling between speed and $\lambda_{\theta}$-stochasticity is lost. 
			The Arnold score $\beta_{\theta}(t)$ (blue trace, lower panel) remains close to $\beta_{\min}=1$,
			affirming $\theta$-wave's quasiperiodicity. Note the antiphasic relationship between the 
			$\beta_{\theta}$-stochasticity and the acceleration $a(t)$ (the latter graph is shifted upwards to
			match the mean level $\langle\beta_{\theta}\rangle$): $\theta$-periodicity loosens as the animal
			slows down ($\beta_{\theta}$-splashes correlate with animal's deceleration) and sharpens as he 
			speeds up. \textbf{E}. Locally averaged $\hat{\lambda}_{\theta}$-score grows with speed, whereas 
			$\hat{\beta}_{\theta}$ tends to drop down with acceleration.
	}}
	\label{fig:thstoch}
\end{figure}

\textbf{1. $\theta$-waves} ($4-12$ Hz, \cite{Burgess,BuzTheta1}) are known to correlate with the animal's motion
state, which suggests that the sample sequences should be selected at a behavioral scale \cite{BuzTheta2,Richard,
Kropff}. In the analyzed experiments, the mice shuttled between two food wells on a U-shaped track, spending
about $22$ secs per lap (average for $5$ mice, for both inbound and outbound runs) and consumed food reward 
over $17$ secs (Fig.~\ref{fig:thstoch}A). On the other hand, the intervals between successive $\theta$-peaks 
distribute around the characteristic $\theta$-period, $\overline{T}_{\theta}\approx 1/m_{\theta}\approx 110$ 
msecs, which defines the timescale of oscillatory dynamics (Fig.~\ref{fig:thstoch}A). To accommodate both 
timescales, we used periods required to complete $1/6^\textrm{th}$ of the run between the food wells, $L_{\theta}
\approx 3.6$ secs, containing about $20-30$ peaks---large enough to produce stable $\lambda$- and $\beta$- scores
\cite{Bol1,Vrbik1,Vrbik2}, but short enough to capture the ongoing dynamics of $\theta$-patterns.

The resulting mean Kolmogorov score $\langle\lambda_{\theta}\rangle=0.54\pm0.12$ is low, indicating
that, on average, $\theta$-cycles closely follow the prescribed trend (\ref{lin}). The mean Arnold score 
$\langle \beta_{\theta} \rangle = 1.1\pm 0.03\approx\beta^{\min}$ also points at the near-periodic behavior 
of the $\theta$-wave (Fig.~\ref{fig:thstoch}C). Nonetheless, $\theta$-patterns exhibit complex dynamics that
differ between quiescence and active states and couple to the animal's speed and acceleration.

\textit{Fast moves}. As mentioned above, the experimental design enforces recurrent behavior, in  which speed
goes up and down repeatedly as the animal moves between the food wells (Fig.~\ref{fig:thstoch}B). When the mouse
moves methodically (lap time less than $25$ sec), $\lambda_{\theta}$ rises and falls along with the speed with
surprising persistence (Fig.~\ref{fig:thstoch}D). Yet, the $\theta$-patterns appearing in this process are 
stochastically generic---the entire sequence of $\lambda_{\theta}$-values remains mostly within the ``domain
of stochastic typicality'' (pink stripe on Figs.~\ref{fig:thstoch}C, D and \ref{fig:stochs}B), below the 
universal mean $\lambda^{\ast}$. However, the patterns become overtly structured as the animal slows 
down, when the Kolmogorov scores drop below $\lambda_{\theta}\approx 0.1$, exhibiting uncommon compliance with
the mean behavior. Such values of $\lambda_{\theta}$ can occur by chance with vanishingly small probability 
$\Phi(0.1)<10^{-17}$ (see formula (\ref{sml}) in Sec.~\ref{sec:met}), which, together with small
$\beta_{\theta}$-score, $\beta_{\theta}\approx 1$, imply that limited motor driving reduces $\theta$-wave 
to a simple nearly-harmonic oscillation with a base frequency $\nu\approx 8$ Hz.

Increasing speed randomizes $\theta$-patterns: the faster the mouse moves, the higher the 
$\lambda_{\theta}$-score. Furthermore, the shape of $\lambda_{\theta}(t)$ dependence exhibits an uncanny 
resemblance to the speed profile
$s(t)$ (Fig.~\ref{fig:thstoch}D). To quantify this effect, we used the Dynamic Time Warping (DTW) technique 
that uses a series of local stretches to match two functions---in this case, $\lambda_{\theta}(t)$ and 
$s(t)$---so that the net stretch can be interpreted as separation, or distance\footnote{DTW separation typically
	satisfies the triangle inequality, $D(a,b)+D(b,c)\geq D(a,c)$, which permits interpreting it geometrically,
	as a distance between signals \cite{Neamtu}.} between functions in ``feature space" \cite{Berndt,Salvador}.
In our case, the DTW-distance between the speed $s(t)$ and Kolmogorov score $\lambda_{\theta}(t)$ during active
moves is small, $D(\lambda_{\theta},s)=19.6\%$, indicating that the deviations of the $\theta$-patterns
from the mean reflect the animals' mobility (SFig.~1).

Note that DTW-affinity between $\lambda_{\theta}(t)$ and speed does not necessarily imply a direct functional
dependence between these quantities. Indeed, plotting points with coordinates $(s,\lambda_{\theta})$ yields 
scattered clouds, suggesting a broad trend, rather than a strict relationship (Fig.~\ref{fig:thstoch}E).
However, if the $\lambda$-scores and the speeds are \textit{locally averaged}, i.e., if each individual $s$- 
and $\lambda$-value is replaced by the mean of itself and its adjacents, then the pairs of such \textit{local
	means}, $(\hat{s}_i,\hat{\lambda}_i)$, reveal a core dependence: increasing speed of the animal entails
higher variability of the $\theta$-patterns. 

In the meantime, the Arnold stochasticity score, $\beta_{\theta}(t)$, is closely correlated with the mouse's
acceleration, $a(t)$. As shown on Fig.~\ref{fig:thstoch}D, the $\beta_{\theta}$-score rises as the mouse 
decelerates ($\theta$-wave clumps) and falls when he accelerates ($\theta$-wave becomes more orderly),
producing a curious antiphasic $\beta_{\theta}\textrm{-}a$ relationship, which is also captured by the local
averages $(\hat{a}_i,\hat{\beta}_i)$ (Fig.~\ref{fig:thstoch}E). The distance between $\beta_{\theta}(t)$
and $-a(t)$ (the minus sign accounts for the antiphase) is $D(\beta_{\theta},-a)=33.6\%$, which means that
speed influences the $\theta$-wave's statistical typicality more than acceleration impacts its orderliness. 

\textit{Slow moves}. When the mouse meanders and slows down (lapse time over $25$ sec), $\theta$-patterns 
change: the $\lambda_{\theta}$-score increases in magnitude and uncouples from speed, (DTW distance is twice
that of the fast moves case, $D(\lambda_{\theta},s)=38.8\%$), suggesting that, without active motor driving,
$\theta$-rhythmicity is less controlled by the mean oscillatory rate, i.e., is more randomized. The Arnold's 
parameter $\beta_{\theta}$ also slightly increases, $D(\beta_{\theta},-a)=36.1\%$, indicating concomitant 
$\theta$-disorder.

Overall, the combined $\lambda_{\theta}\textrm{-}\beta_{\theta}$ dynamics suggests that during active behavior, 
the shape of the $\theta$-wave is strongly controlled by the mouse's moves. Highly ordered, nearly periodic
$\theta$-peaks appear when the animal starts running---the $\theta$-frequency range then narrows to the mean,
expected value. The increasing speed stirs up the $\theta$-patterns; the disorder grows and reaches its maximum
when the animal moves fastest and begins to slow down. During periods of inactivity, the coupling between 
$\theta$-patterns and speed is weakened and then reinforced as the mouse stiffens his resolve.

We emphasize however, that these dependencies should not be viewed as na\"ive manifestations of known couplings
between instantaneous or time-averaged frequency with the animals' speed or acceleration \cite{Richard,Kropff}. 
Indeed, the $\theta$-frequency alters at the same rate as the $\theta$-amplitude---many times over the span of
each $\theta$-pattern, providing an instantaneous characterization of the wave \cite{Vakman,Rice}. In contrast,
the stochasticity parameters describe the wave form as a single entity (Fig.~\ref{fig:thstoch}A,C).
On the other hand, time-averaging levels out fluctuations and highlights mean trends, whereas Kolmogorov and 
Arnold parameters are sensitive to individual elements in the data sequences. \textit{Thus, $\lambda$ and 
$\beta$ scores describe  of wave shapes without defeaturing, putting each pattern, as a whole, into a 
statistical perspective}. It hence becomes possible to approach questions addressed in the Introduction: 
identify typical and atypical wave patterns, quantify levels of their  orderliness, detect deviations from 
natural behavior and so forth. 
 
\begin{figure}
	\centering
	\includegraphics[scale=0.8]{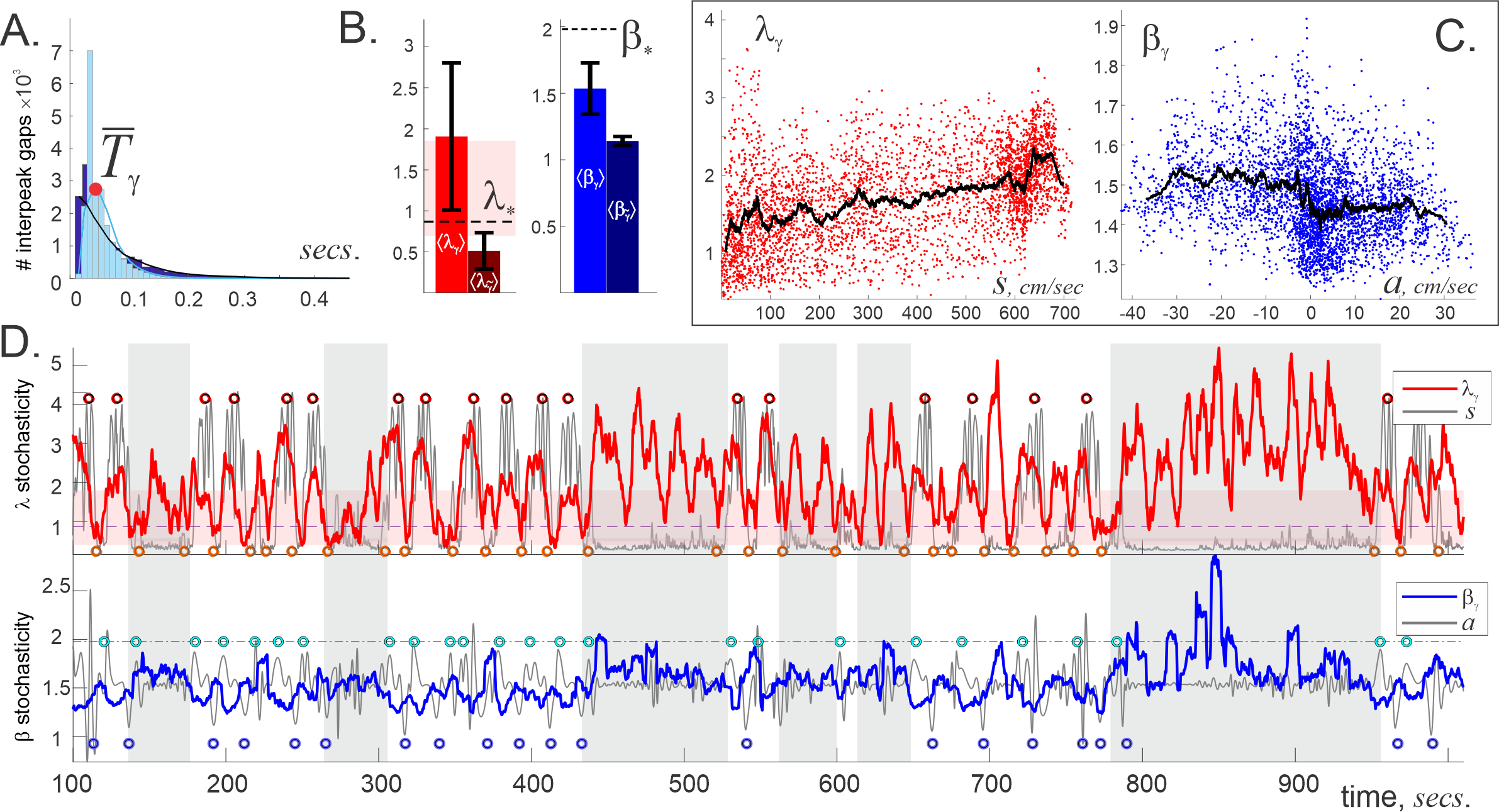}
	\caption{{\footnotesize
			\textbf{$\gamma$-wave stochasticity}. 
			\textbf{A}. A histogram of $\gamma$-interpeak intervals exhibits an exponential-like distribution
			with mean characteristic $\gamma$-period, $\overline{T}_{\gamma}=18.6\pm 1.9$ msec, about six times
			smaller than $\overline{T}_{\theta}$. 
			\textbf{B}. The average scores $\langle\beta_{\gamma}\rangle$ and $\langle\lambda_{\gamma}\rangle$
			are higher than for the $\theta$-wave, indicating that $\gamma$-patterns are more diverse than 
			$\theta$-patterns.
			\textbf{C}. Locally averaged $\hat{\lambda}_{\gamma}$-score grows with speed, while 
			$\hat{\beta}_{\gamma}$ switches from higher to lower level with increasing acceleration. 
			\textbf{D}. The dynamics of the $\lambda_{\gamma}$-score (top panel) correlates with changes in the
			speed when the animal moves actively. Note that $\lambda_{\gamma}$ often exceeds the upper bound of
			the ``pink stripe,'' i.e., $\gamma$-waves often produce statistically uncommon patterns, especially 
			during rapid moves. The $\beta_{\gamma}$-score (bottom panel) correlates with the animal's acceleration,
			which is lost when lap times increase (gray stripes).
	}}
	\label{fig:gstoch}
\end{figure}

\textbf{2. $\gamma$-waves} ($30-80$ Hz, \cite{ColginGm}) exhibit a wider variety of patterns than $\theta$-waves.
The interpeak intervals between consecutive $\gamma$-peaks, $T_{\gamma}$, are nearly-exponentially distributed,
which implies that both smaller and wider $\gamma$-intervals are statistically more common 
(Fig.~\ref{fig:gstoch}A).

\textit{Fast moves}. For consistency, the sample sequences, $X_{\gamma}$, were drawn from the same time windows,
$L_{\gamma}=L_{\theta}\approx 6$ secs, which contained, on average, about $300$ elements that yield a mean
Kolmogorov score $\langle\lambda_{\gamma}\rangle=1.84\pm 1.03$---more than twice higher than the impartial mean
$\lambda^{\ast}$ and three times above the $\langle\lambda_{\theta}\rangle$ score. Such values can randomly 
occur with probability $1-\Phi(1.84)\lesssim 2\cdot 10^{-3}$, which suggests that generic $\gamma$-patterns are
statistically atypical and may hence reflect organized network dynamics, rather than random extracellular field
fluctuations (Fig.~\ref{fig:gstoch}B). The average Arnold parameter also grows compared to the $\theta$-case, 
but remains lower than the impartial mean, $\langle\beta_{\gamma}\rangle=1.61\pm0.53<\beta^{\ast}$, implying 
that, although $\gamma$-waves are more disordered than the $\theta$-waves, they remain overall oscillatory.

On average (for all five mice), the Kolmogorov score, $\lambda_{\gamma}(t)$, escapes the domain of ``stochastic
typicality" approximately half of time through transitions that closely follow speed dynamics. Together with the
observation that $\lambda_\gamma$ takes a larger range of values than $\lambda_\theta$ ($\gamma$-patterns deviate
more from the average as speed increases), this suggests that there is a greater diversity of $\gamma$-responses
to movements (Fig.~\ref{fig:gstoch}C). In particular, the dependence between locally averaged $\hat{\lambda}
_{\gamma}(t)$ and $\hat{s}(t)$ is less tight: the average DTW distance, $D(\lambda_{\gamma},s)\approx 23.4\%$, is
slightly higher than the distance in the $\lambda_{\theta}(t)$ dynamics, which illustrates that 
$\gamma$-patterns are less sensitive to speed than $\theta$-patterns.

From a structural, $\beta$-perspective, $\gamma$-wave becomes closer to periodic when an actively moving animal
slows down: during these periods, $\beta_{\gamma}$-score reduces close to its minimal value, when the 
corresponding Kolmogorov score also drops to $\lambda_{\gamma}\approx0.2$. Since the latter is unlikely to occur
by chance (cumulative probability of that is $\Phi(0.2)\lesssim10^{-12}$), these changes may represent structured
network dynamics. The highest deviations of $\gamma$-patterns from the mean ($\lambda_{\gamma}\gtrsim 3$) are 
accompanied by high $\beta_{\gamma}$-scores, which happen as the mice slow down from maximal speed and implies
that circuit activity is least structured during these periods (Fig.~\ref{fig:gstoch}D).
The relation between $\gamma$-orderliness, $\beta_{\gamma}(t)$, and acceleration is also similar to the 
corresponding dependence in the $\theta$-case: acceleration induces stricter $\gamma$-rhythmicity and
deceleration randomizes $\gamma$-patterns, with about the same overall DTW distance, $D(\beta_{\gamma},-a)\approx
34.4\%$. 

\textit{During slower movements}, the $\gamma$-dynamics change qualitatively: the magnitudes of both 
$\lambda_{\gamma}(t)$ and $\beta_{\gamma}(t)$ grow higher, indicating that decoupling from motor activity
enforces statistically atypical $\gamma$-rhythmicity in the hippocampal network, as in the $\theta$-waves. 
In particular, the uncommonly high $\beta_{\gamma}$ scores point at frequent $\gamma$-bursting during quiescence. 

Once again, we emphasize that these results do not represent known correlations between instantaneous or 
time-averaged $\gamma$-characteristics and motion parameters \cite{Ahmed,Montgomery}. Rather, the outlined 
$\lambda_{\gamma}(t)$ and $\beta_{\gamma}(t)$ dependencies capture pattern-level dynamics of $\gamma$-waves
that reflect circuit activity at an intermediate timescale. As an illustration, note that the amplitude of
$\gamma$-waves, $\tilde{\gamma}$, along with the instantaneous $\gamma$-frequency, $\omega_{\gamma}$, have low
stochasticity scores, comparable to the ones produced by the Poisson process, $\langle\beta_{\tilde{\gamma}}
\rangle= 1.15\pm 0.08$ and $\langle\lambda_{\tilde{\gamma}}\rangle=0.52\pm 0.26$ (Fig.~\ref{fig:rndstoch}).
Thus, although instantaneous characteristics exhibit restrained, quasiperiodic behavior, they allow a rich 
morphological variety of the underlying $\gamma$-oscillations.

\begin{figure}
	\centering
	\includegraphics[scale=0.8]{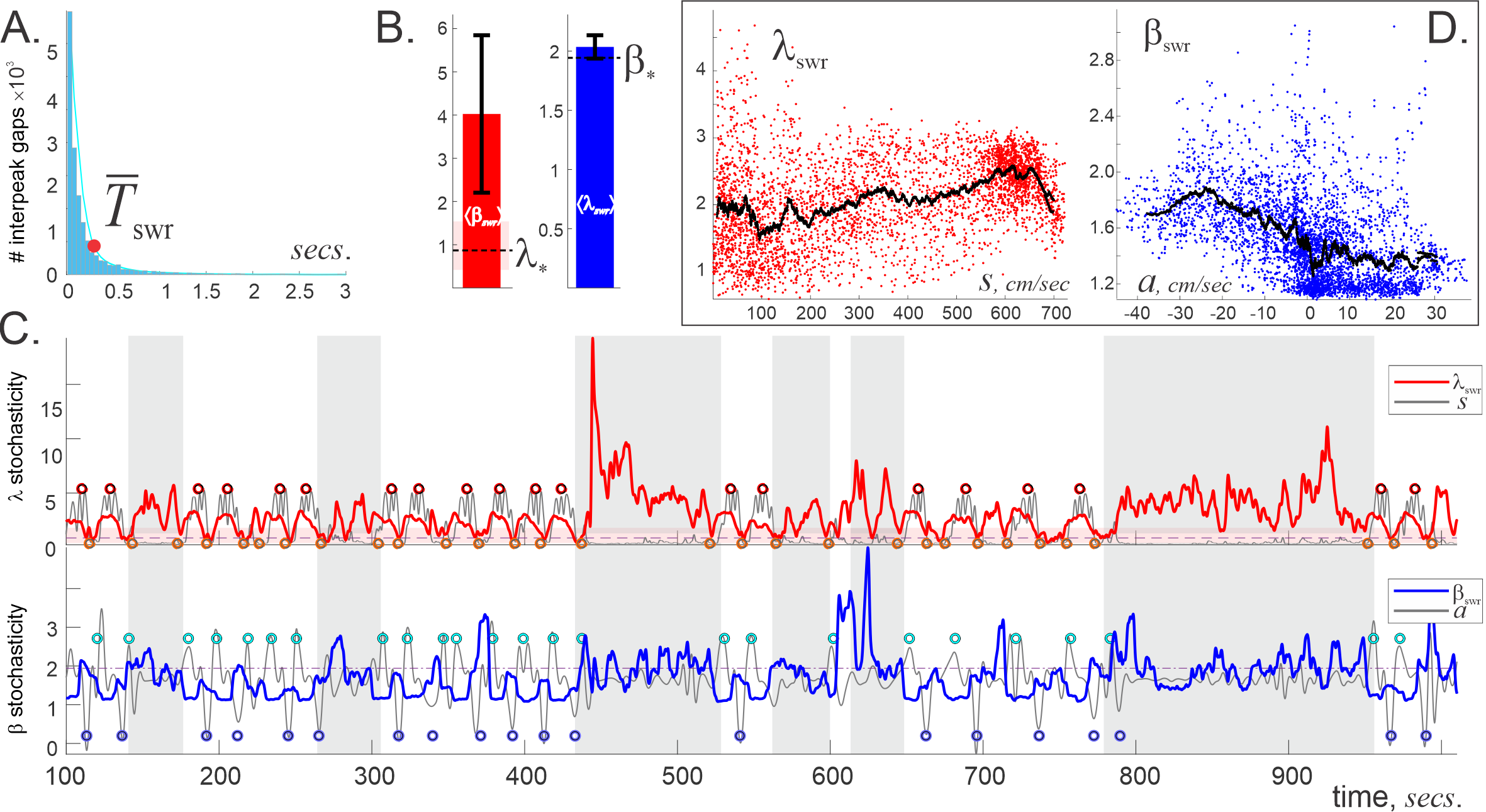}
	\caption{{\footnotesize
			\textbf{Sharp Wave-Ripples' stochasticity}. 
			\textbf{A}. A histogram of intervals between SWR events is nearly exponential. 
			\textbf{B}. The averages $\langle\lambda_{\swr}\rangle$ and $\langle\beta_{\swr}\rangle$ are high, 
			indicating both frequent deviation of SWR events from the mean and higher temporal clustering than
			for the $\theta$ and $\gamma$-patterns. 
			\textbf{C}. The animal's speed (gray line, top panel) correlates with the Kolmogorov parameter 
			$\lambda_{\swr}$ during fast exploratory lapses. During inactivity (vertical gray stripes) the 
			$\lambda_{\swr}$-stochasticity uncouples from speed, exhibiting high spikes that mark strong
			``fibrillation'' of SWR patterns. The antiphasic relationship between the animal's acceleration
			$a(t)$ (gray line, bottom panel) and Arnold score $\beta_{\swr}(t)$ shows that SWRs tend to cluster
			when as the animal decelerates, while acceleration enforces periodicity. During slower moves (gray
			stripes), the relationship between speed, acceleration, and stochasticity is washed out and stochastically
			improbable patterns dominate.
			\textbf{D}. Locally averaged $\hat{\lambda}_{\swr}$ grows with speed and $\hat{\beta}_{\swr}$ drops
			with acceleration.
	}}
	\label{fig:swr}
\end{figure} 

\textbf{3. Sharp Wave Ripples} (SWRs), the high amplitude splashes (over $2-3$ standard deviations from the 
mean) of high frequency waves  ($150-250$ Hz, \cite{ColginR}), exhibit the richest pattern dynamics.

\textit{During fast moves}, SWR events appear at approximately the same exponential rate as the 
$\tilde{\gamma}$-peaks, $\overline{T}_{\swr}\approx \overline{T}_{\tilde{\gamma}}$ msec, but exhibit higher
$\lambda$-scores, $\langle \lambda_{\swr}\rangle=2.40\pm1.57$, over the same sampling periods $L_{\swr}\approx
6$ sec (Fig.~\ref{fig:swr}A,B). The low probability of these patterns ($1-\Phi(2.5)\lesssim 10^{-6}$) and the
relatively high mean $\beta$-score, $\langle\beta_{\swr}\rangle=1.71\pm0.64$, indicate that SWRs tend to 
exhibit intermittent clustering that may reflect brisk, time-localized circuit activity, such as rapid replays
and preplays of the hippocampal place cells \cite{Girardeau2,Singer,Roux,Sadowski2,Denovellis,BarneSpars,Wu}.

Interestingly, SWR-patterns also correlate with the animal's speed profile about as much as $\gamma$-patterns,
$D(\lambda_{\swr},s)\approx 23.8\%$ \cite{Denovellis}. The $\beta_{\swr}(t)$-dependence displays the familiar
antiphasic relationship with the animal's acceleration---SWR events tend to cluster more when the animal slows
down (Fig.~\ref{fig:swr}C,D). However, orderliness of SWRs is driven by acceleration stronger than orderliness
of $\gamma$-patterns: the range of $\beta_{\swr}$-scores is twice as wide as the range of $\beta_{\gamma}$-scores
(broader pattern variety), with a similar DTW distance $D(\beta_{\swr},-a)\approx39.7\%$. 

\textit{During quiescent periods}, both $\lambda_{\swr}$ and $\beta_{\swr}$ grow and exhibit extremely high
spikes, indicating that endogenous network dynamics produce stochastically improbable, highly clustered SWR
sequences. Physiologically, these statistically uncommon SWR patterns may indicate sleep or still wakefulness
replay activity, known to play an important role in memory consolidation \cite{Kudrimoti,ONeil}.

Overall, the temporal clumping comes forth as a characteristic feature of the SWR events, suggesting that SWRs
are manifestations of fast, targeted network dynamics that brusquely ``ripple" the extracellular field, unlike
the rhythmic $\theta$ and $\gamma$-undulations \cite{ColginR}.

\subsection{Stochasticity in space}
\label{sec:space}

\begin{figure}[h]
	\centering
	\includegraphics[scale=0.79]{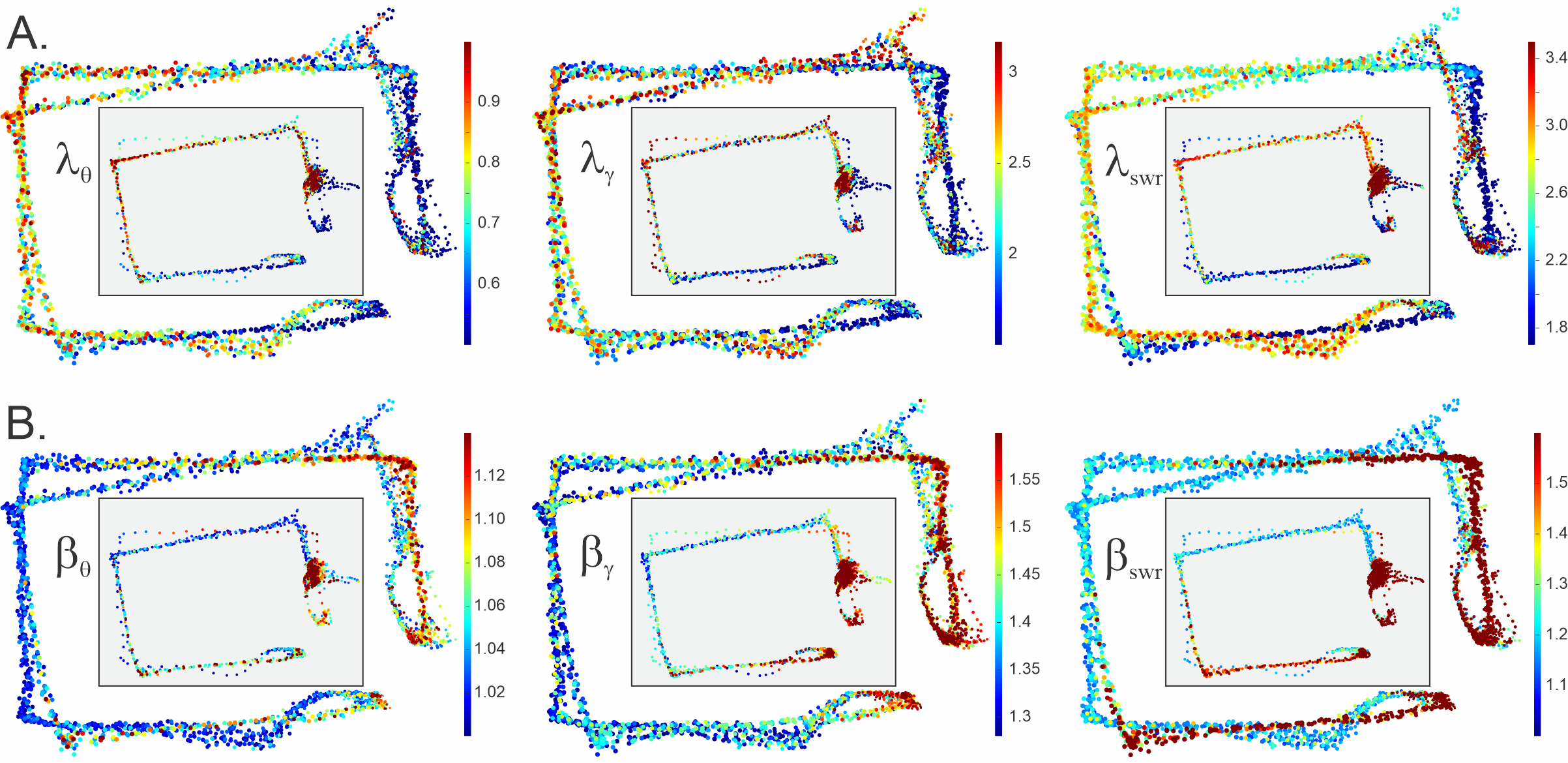}
	\caption{{\footnotesize
			\textbf{Spatial stochasticity maps} were obtained by plotting $\lambda$ and $\beta$ parameters along
			the	trajectory.
			\textbf{A}. The $\lambda$-maps show that $\theta$-wave, $\gamma$-wave and SWRs generally follow the
			mean trend near the food wells (with scattered wisps of high stochasticity) and deviate from the mean
			mostly over the areas most distant from the food wells. The smaller maps in 
			the gray boxes represent slow lapses: the overall layout of high-$\lambda$ and low-$\lambda$ fields
			is same as during the fast moves, which suggests spatiality of $\lambda$-stochasticity.
			\textbf{B}. The behavior of $\beta_{\theta}$ is opposite: the ``uneventful,'' distant run segments 
			attract nearly-periodic behavior, while the food wells attract time-clumping wave patterns.
	}}
	\label{fig:space}
\end{figure}

Distributing the $\lambda$ and $\beta$ scores along the animal's trajectory yields \textit{spatial maps of
stochasticity} for each brain rhythm and reveals a curious spatial organization of LFP patterns with similar
morphology. As shown on Fig.~\ref{fig:space}, higher $\lambda$-values for all waves are attracted to segments
where the mouse is actively running with maximal speed, furthest away from the food wells. Patterns that are 
close to the expected average (low-$\lambda$) concentrate in the vicinity of food wells where the animal moves
slowly. The latter domains also tend to host high $\beta$-scores that appear as the animal approaches the food
wells, as well as the lowest $\beta$s, which appear as the animal accelerates away \cite{BarneSpars}. In other
words, the LFP waves become more ``trendy'' and, at the same time, more structured (either more periodic or more
clustered) over the behaviorally important places (e.g., food wells) that require higher cognitive activity. 
On the other hand, the outer parts of the track, where the brain waves are less controlled by the mean and 
remain moderately disordered, are marked by uncommon patterns.

Intriguingly, the same map structure is reproduced during slow lapses, when the motor control of the patterns
weakens, suggesting that speed and acceleration are not the only determinants of the LFP patterns. As shown 
on Fig.~\ref{fig:space}, even when the mouse dawdles, the waves tend to deviate from the mean around the outer 
corners and follow the mean in the vicinity of the food wells. Similarly, the patterns start clumping as the
mouse approaches the food wells, and distribute more evenly as he moves away.

These results suggest that spatial context may, by itself, influence hippocampal brain rhythm structure, which
is reminiscent of the place-specific activity exhibited by spatially tuned neurons, e.g., place cells 
\cite{MosRev} or parietal neurons \cite{Nitz1}. For example, the ``bursting'' (high-$\beta$) fields and 
``domains of evenness" (small $\lambda$) surround food wells; the quasiperiodicity fields (small $\beta$s) 
as well as ``wobbling-waves'' (large $\lambda$s) stretch over the outer segments (Fig.~\ref{fig:space}). 
Physiologically, this ``spatiality of stochasticity" may reflect a coupling between the hippocampal
place-specific spiking activity and extracellular field oscillations. 

\subsection{$\lambda$-$\beta$ relationships}
\label{sec:lb}

\begin{figure}[h]
	\centering
	\includegraphics[scale=0.8]{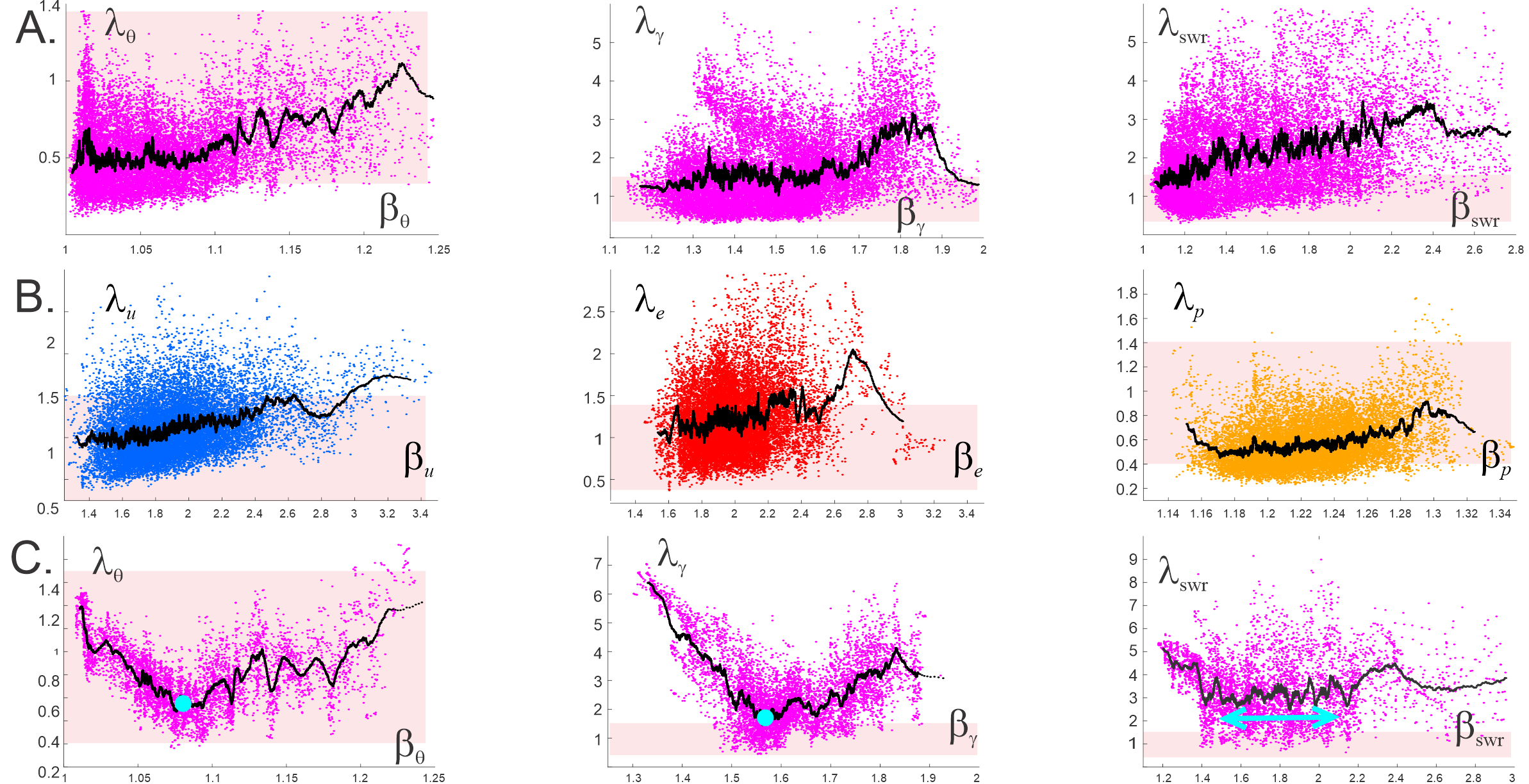}
	\caption{{\footnotesize
			\textbf{Dependencies between stochasticity parameters}. 
			\textbf{A}. Points with coordinates $(\beta_i,\lambda_i)$ computed for each individual sample 
			sequence produce clouds that imply no strict $\beta(\lambda)$ dependence. However, locally 
			averaged stochasticity parameters $(\hat{\lambda},\hat{\beta})$ exhibit much tighter 
			relationships (black dots). The growing $\hat{\beta}(\hat{\lambda})$ trends indicate that ordered,
			semiperiodic sequences (low $\hat{\beta}$) tend to accompany samples that comply with the expected
			behavior (low $\hat{\lambda}$). Conversely, patterns tend to fibrillate more as they deviate farther
			from the mean and abandon the pink stripe of ``stochastic typicality.''
			\textbf{B}. The $(\beta,\lambda)$ pairs for sequences drawn from the three random distributions 
			(Fig.~\ref{fig:rndstoch}) produce similar clouds (uniform, blue dots; exponential, red dots; 
			Poisson, orange dots).  
			Locally averaged scores indicate tight $(\hat{\beta}\textrm{-}\hat{\lambda})$ relations  for all
			these series trend similarly to the brain waves' local averages.
			\textbf{C}. The $\hat{\beta}(\hat{\lambda})$ trends evaluated for individual mice may exhibit
			non-generic features. In this case, the oscillatory rate of semi-periodic $\theta$-, $\gamma$- and
			SWR sequences (low $\hat{\beta}$s) deviates significantly from the predicted mean (large 
			$\hat{\lambda}$s). As the disorder increases, the oscillatory rate gets closer to the predicted mean,
			reaching the minimal $(\hat{\beta}_{\theta}\textrm{-}\hat{\lambda}_{\theta})$ and 
			$(\hat{\beta}_{\gamma}\textrm{-}\hat{\lambda}_{\gamma})$ combinations (cyan dot) before retaking a
			joint growth trend. In case of the SWRs, the minimum is spread into a plateau where changes in
			$\hat{\beta}_{\swr}$ do not affect $\hat{\lambda}_{\swr}$.
	}}
	\label{fig:bl}
\end{figure}

The definitions of the stochasticity scores $\beta$ and $\lambda$ do not imply an \textit{a priori} 
$\beta\textrm{-}\lambda$ relationship. Indeed, plotting sets of points with coordinates $(\beta,\lambda)$ 
for all sample sequences, $X_{\theta}$, $X_{\gamma}$ and $X_{\swr}$, yields scattered clouds, rather than
curve-like graphs (Fig.~\ref{fig:bl}A). However, computational studies of number-theoretic sequences carried
in \cite{ArnoldB1,ArnoldB2,ArnoldB3,ArnoldB4} suggest that such behavior may be caused by the occasional large
contributions from atypical sequences and that local smoothing may yield much tighter couplings between 
the stochasticity parameters.

To study whether similar phenomena take place in the LFP sequences, we evaluated the the local averages
$(\hat{\beta}_i,\hat{\lambda}_i)$, which revealed $\hat{\beta}\textrm{-}\hat{\lambda}$ relationships
illustrated on Fig.~\ref{fig:bl}. For the full dataset that includes all mice, we observe that orderly 
sequences (low $\hat{\beta}$-scores) tend to follow the prescribed mean trend more closely (low $\hat{
	\lambda}$s). In contrast, patterns that deviate from the expected mean tend to produce disordered
and clumping sequences, notably for $\gamma$-waves and SWRs.

Curiously, similar $\hat{\beta}\textrm{-}\hat{\lambda}$ behavior are exhibited by random sequences (uniformly,
exponentially and Poisson-distributed, Fig.~\ref{fig:bl}B), which suggests that the tendency of $\lambda$ to
rise with growing $\beta$, observed in large volumes of heterogeneous data (different mice) may not be of a 
specifically physiological nature, but may reflect mathematical connections between the 
stochasticity indices \cite{Arnold3,Arnold4,Arnold5}.

In view of these results, it is surprising that individual mice can exhibit personalized dependencies between
predictability, $\lambda$, and orderliness, $\beta$, of their LFP patterns. In the case illustrated on 
Fig.~\ref{fig:bl}, nearly periodic (small $\hat{\beta}_{\theta}$) $\theta$-patterns tend to deviate from the 
predicted behavior quite significantly. As the disorder increases, the patterns become more compliant with the 
underlying mean, until a trough of $\hat{\beta}_{\theta}\textrm{-}\hat{\lambda}_{\theta}$ dependence is reached.
Then the tendency is reversed: growing $\theta$-disorder is accompanied by further deviation from the mean. Note
that the entire stochasticity dynamics remain within the ``typicality zone," $0.4\leq\hat{\lambda}_{\theta}\leq
1.8$. Analogous behavior is exhibited by the $\gamma$-wave but at a larger scale: orderly $\gamma$-waves, 
$\hat{\beta}_{\gamma}\approx1.4$, are concomitant with $\lambda$-scores as high as $\lambda_{\gamma}\approx6.5$ 
(highly improbable patterns, $1-\Phi(6.5)\lesssim 10^{-50}$). As $\hat{\beta}_{\gamma}$ grows, $\hat{\lambda}_{\gamma}$
decreases, approaching stochastic commonality at the lowest point. Then, as $\hat{\beta}_{\gamma}$ grows further,
the disordered sequences increasingly deviate from the expected mean. The $\hat{\beta}_{\swr}\textrm{-}\hat{\lambda}_{\swr}$
dependence is less tight but exhibits a similar trend: at first, small $\hat{\beta}_{\swr}$'s pair with higher
$\hat{\lambda}_{\swr}$'s, then level out over an intermediate range of $\beta$'s (the minimum is flattened out
in contrast with $\theta$- and $\gamma$-waves), and then grows again with the increasing SWR-disorder.

Importantly, we found no relationship between the $\lambda_{\theta}$, $\lambda_{\gamma}$, and $\lambda_{swr}$,
nor between $\beta_{\theta}$, $\beta_{\gamma}$, and $\beta_{swr}$ (SFig.~2), which implies that the $\lambda_
{\ast}$ and the $\beta_{\ast}$-scores associated with different waves are largely independent, providing their
own, autonomous characterizations of the wave shapes.

\section{Discussion}

The recorded LFP signals are superpositions of locally induced extracellular fields and inputs transmitted from
anatomically remote networks. The undulatory appearance of the LFP is often interpreted as a sign of structural
and functional regularity\footnote{A succinct expression of this view is provided in \cite{Fransen}:
	\textit{``rhythmicity is the extent to which future phases can be predicted from the present one.''}}, 
but the dynamics of these oscillations is actually highly complex. Understanding the balance between 
deterministic and stochastic components in LFPs, as well as questions about their continuity and discreteness
pose significant conceptual challenges, as it happened previously in other disciplines\footnote{In his 1955 
discussion of the foundations of Quantum Mechanics, John von Neumann attributes a great significance to the
fact that \textit{``...the general opinion in theoretical physics had accepted the idea that ...continuity
		...is merely simulated by an averaging process in a world which in truth discontinuous by its very 
		nature. This simulation is such that man generally perceives the sum of many billions of elementary 
		processes simultaneously, so that the leveling law of large numbers completely obscures the real 
		nature of the individual processes.''} \cite{Neumann}}.

Structurally, LFP rhythms may be described through discrete sequences of wave features (heights of peaks, 
specific phases, interpeak intervals, etc.), or viewed as transient series---bursts---of events, as in the
case of SWRs \cite{Ede}. It is well recognized that such sequences are hard to decipher and to forecast, e.g.,
a recent discussion of a possible role of bursts in brain waves' genesis posits: \textit{``An important feature
	that 
sets the burst scenario apart is the lack of continuous phase-progression between successive time points---and
therefore the ability to predict the future phase of the signal---at least beyond the borders of individual bursts"}
\cite{Ede}. In other words, the nonlinearity of LFP dynamics, as well as its transience and sporadic external
driving, result in effective stochasticity of LFP patterns---an observation that opens a new round of inquiries 
\cite{Ede,Jones}. 
For example, how exactly should one interpret the ``unpredictability'' of a temporal sequence? Does it mean that
its pattern cannot be resolved by a particular algorithm, or that it is unpredictable in principle, ``genuinely
random," such as a gambling sequence? What is the difference between the two? How is the apparent randomness of
LFP rhythms manifested physiologically? Are the actual network computations based on ``overcoming the apparent 
randomness" and somehow deriving the upcoming phases or amplitudes from the preceding ones, or may there be 
alternative ways of extracting information? Does the result depend on the ``degree of randomness" and if so, 
then how to distinguish between a ``more random'' and a ``less random'' patterns? These questions are not 
technical, pertaining to a specific mechanism, nor specifically neurophysiological; rather, these are 
fundamental problems that transcend the field of neuroscience. Historically, similar questions have motivated
mathematical definitions of randomness that are still debated to this day \cite{Mises,Uspenskii,Volchan}. 

One approach to addressing these issues was suggested by Kolmogorov in 1933 (also the year when brain waves
were discovered \cite{Berger1}), based on the statistical universality of stochastic deviations from the 
expected behavior \cite{Kolmogorov,Stephens}. From Kolmogorov's perspective, randomness is contextual: if a
sequence $X$ deviates from an expected mean behavior within bounds established by the distribution (\ref{phi}),
then $X$ is \textit{effectively} random. In other words, an individual sequence may be viewed as random if it
could be randomly drawn from a large pool of similarly trending sequences, with sufficiently high probability.
This view permits an important conceptual relativism: even if a sequence is produced by a specific mechanism
or algorithm, it can still be viewed as random as long as its $\lambda$-score is ``typical" according to the
statistics (\ref{phi}). For example, it can be argued that geometric sequences are typically more random than
arithmetic ones, although both are defined by explicit formulae \cite{Arnold1,Arnold2,Arnold3,Arnold4,Arnold5}.
By analogy, individual sample sequences of $\theta$, $\gamma$, or SWR events may be generated by specific 
synchronization mechanisms at a precise timescale, and yet they may be empirically classified and quantified
as stochastic.

A practical advantage of Kolmogorov's approach is that mean trends, such as (\ref{lin}), can often be reliably
established, interpreted, and then used for putting the stochasticity of the underlying sample patterns into a
statistical perspective. Correspondingly, assessments based on $\lambda$-scores were previously applied in a 
variety of disciplines from genetics \cite{KolMen,Stark,Gurzadyan1} to astronomy \cite{Gurzadyan2}, and from 
economics \cite{Brandouy} to number theory \cite{Arnold2,Arnold3,Arnold1,Arnold4,Arnold5,ArnoldB1,ArnoldB2,ArnoldB3,
	ArnoldB4,Christoph,Ford}. Some work has also been done in brain wave analyses, e.g., for testing normality 
of electroencephalograms' long-term statistics \cite{Weiss1,Weiss2,Weiss3,McEwen}. Arnold's $\beta$-score 
provides an independent assessment of orderliness (whether elements of an arrangement tend to attract, repel
or be independent of each other) and it has not been, to our knowledge, previously used in applications.

Shifting window analyses ground the $\lambda$- and $\beta$-values in the context of preceding and upcoming
observations. Since neighboring patterns change only marginally, the time-dependent $\lambda(t)$ and 
$\beta(t)$ describe quasi-continuous pattern dynamics at a \textit{mesoscale}---over the span of several 
undulations---which complements the microscale (instantaneous) and macroscale (time-averaged) assessments. 
While the individual, ``stroboscopically selected" patterns can be viewed as stochastic, the continuous 
$\lambda(t)$ and $\beta(t)$ dependencies describe ongoing pattern dynamics. 

Importantly, Kolmogorov's and Arnold's scores are impartial and independent from physiological specifics
or contexts, thus providing self-contained semantics for describing the LFP data and a novel venue for
analyzing the underlying neuronal mechanisms. It becomes possible to distinguish ``statistically mundane" LFP
patterns from exceptional ones and to capture the transitions between them, as well as to link pattern dynamics
to changes in the underlying network's dynamics (Figs.~\ref{fig:thstoch}-\ref{fig:space}). For example, since
$\theta$-bursts are physiologically linked to long-term synaptic potentiation \cite{Greenstein,Hinder}, 
$\theta$-patterns with high $\beta$-scores may serve as markers of plasticity processes taking place in the
hippocampal network at specific times and places \cite{Larson,Sheridan,Nguyen}. 
Furthermore, high-$\beta_{\theta}$ regions near food wells indicate that reward proximity may trigger
hippocampal plasticity and, since hippocampal neurons' spiking is coupled to $\theta$-cycles \cite{Skaggs}, 
have a particular effect on memory (Fig.~\ref{fig:space}). On the other hand, low-$\beta_{\theta}$ indicates
limit cycles in the network's phase space that uphold simple
periodicity. $\gamma$-bursts (high $\beta_{\gamma}$) mark heightened attention and learning periods 
\cite{ClgMsr,Lundqvist}. In our observations, they appear during the mouse's approach to the reward locations
and disappear as it ventures away from them (Fig.~\ref{fig:space}). Clustering SWR events reflect dense replay
activity \cite{Roux,Singer}, indicative of periods of memory encoding, retrieval, and network 
restructurings \cite{Replays,Sadowski2}. 

Overall, the proposed approach allows studying brain rhythms from a new perspective that complements existing
methodology, which may lead to a deeper understanding of the synchronized neuronal dynamics and its physiological
functions at temporal mesoscale.

\vspace{7pt}

\textbf{Acknowledgments}. We are grateful to Dr. A. Babichev for fruitful discussions. 
C.H. and Y.D. are supported by NIH grant R01NS110806 and NSF grant 1901338.
C.J. and D.J. are supported by NIH grants R01MH112523 and R01NS097764.


\newpage
\section{Mathematical Supplement}
\label{sec:met}

\textbf{Computational algorithms}. 

\begin{wrapfigure}{c}{0.4\textwidth}
	\centering
	\includegraphics[scale=0.8]{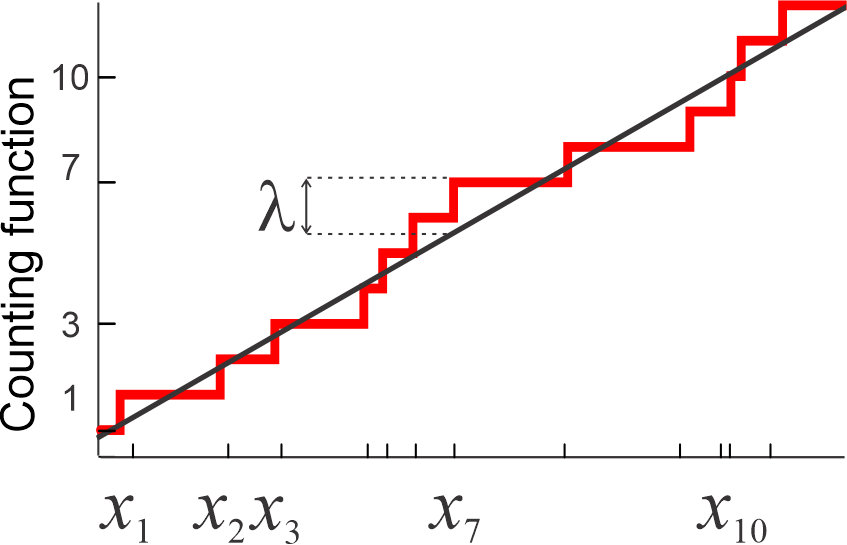}
	\caption{{\footnotesize
			\textbf{Counting function} $N(X)$ (red staircase) makes unit steps at each point of a sequence
			$X=\{x_1,x_2,\ldots,x_n\}$ (tick marks on the $x$-axis). The normalized maximal deviations 
			$\lambda(X)$ from the expected mean $\bar{N}(x)$ (straight line) exhibit statistical universality
			and can hence be used for characterizing stochasticity	of the individual data sequences $X$.
	}}
	\label{fig:stair}
	\end{wrapfigure}

\textit{1. Kolmogorov score}. Let $X={x_1\leq x_2\leq\ldots\leq x_n}$, be an ordered sequence and let
$N(X,L)$ be the number of elements smaller than $L$,
\begin{equation*}
	N(X,L) = \{\textrm{number of} \,\, 0\leq x_k < L\}.
	\label{N}
\end{equation*}
Let $\bar{N}(X,L)$ be the expected number of elements that interval (Fig.~\ref{fig:stair}). The closer $X$
follows the prescribed behavior, the smaller the normalized deviation\footnote{The supremum, rather	than 
maximum, is required in formula (\ref{lam}) due to discontinuity of the counting function $N(X,L)$ at the
stepping points.}
\begin{equation}
	\lambda(X)= \sup_L|N(X,L) - \bar{N}(X,L)|/\sqrt{n}.
	\label{lam}
\end{equation}

A remarkable observation made in \cite{Kolmogorov} is that the cumulative probability of having $\lambda(X)$ 
smaller than a given $\lambda$ converges to the function 
\begin{equation}
	\Phi(\lambda) = \sum_{k=-\infty}^{\infty}(-1)^k e^{-2k^2 \lambda^2},
	\label{phi}
\end{equation}
that starts at $\Phi(0)=0$ and grows to $\Phi(\infty)=1$. The derivative of the cumulative density (\ref{phi})
defines the probability distribution for $\lambda$, $P(\lambda)=\partial_{\lambda}\Phi(\lambda)$
(Fig.~\ref{fig:stochs}B). Even though the range of $P(\lambda)$ includes arbitrarily small or large $\lambda$s,
the shape of the distribution implies that excessively high or low $\lambda$-values are rare, e.g., sequences 
with $\lambda(X)\leq 0.4$ or $\lambda(X)\geq 1.8$ appear with probability less than $0.3\%$, $\Phi(0.4)\approx
0.003$ and $\Phi(1.8)\approx0.997$. Since these statistics are universal, i.e., apply to any sequence $X$, the
$\lambda$-score can serve as a universal measure of ``stochastic typicality" of a pattern
 \cite{Stephens,Arnold1,Arnold2,Arnold3,Arnold4,Arnold5}.

\textit{2. Corrections to Kolmogorov score} up to the order $n^{-3/2}$,
\begin{equation}
\lambda(X)\to\lambda(X)\left(1+\frac{1}{4n}\right)+\frac{1}{6n}-\frac{1}{4n^{3/2}},
\label{lamn}
\end{equation}
allows for an increase in the accuracy of the finite-sample estimates to over $0.01\%$ for sequences containing
as little as $10$-$20$ elements \cite{Bol1,Vrbik1,Vrbik2}. In this study, all $\lambda$-evaluations are based
on the expression (\ref{lamn})  and use data sequences that contain more than $25$ elements. 

\textit{3. Mean Kolmogorov stochasticity score}. The mean $\lambda$ can then be computed as
\begin{equation*}
	\lambda^{\ast}=\int_0^{\infty} \lambda P(\lambda)d\lambda=\int_0^{\infty} \Phi(\lambda)d\lambda,
\end{equation*}
where we used integration by parts and the fact that the distribution $P(\lambda)$ starts at $0$, $P(0)=0$, 
and approaches $0$ at infinity, $P(\infty)=0$ (Fig.~\ref{fig:stochs}B). 
Integrating the Gaussian terms in expansion (\ref{phi}) yields Mercator series
\begin{equation*}
	\lambda^{\ast}=\sqrt{\frac{\pi}{2}}\sum_{k=1}^{\infty}(-1)^{k+1}\frac{1}{k}=\sqrt{\frac{\pi}{2}}\ln2
	\approx 0.8687.
\end{equation*}

\textit{4. $\Phi(\lambda)$ estimates}. For small $\lambda$s, the Kolmogorov's $\Phi$-function (\ref{phi}) can
be approximated by 
\begin{equation}
	\Phi(\lambda)\approx\frac{\sqrt{2\pi}}{\lambda}e^{-\pi^2/8\lambda^2},
	\label{sml}
\end{equation} 
and for large $\lambda$s, it is approximated by the two lowest-order terms in (\ref{phi}), $\Phi(\lambda)
\approx 1-2 e^{-2\lambda^2}$ \cite{Kolmogorov,Stephens}. These formulae allow quick evaluations of the 
$\lambda$-scores' cumulative probabilities outside of the ``stochastic typicality band," $\lambda<0.4$ or
$\lambda>1.8$.

\textit{5. Arnold score}. Let us arrange the points of the sequence $X$ on a circle of length $L$ and consider
the arcs between pairs of consecutive elements, $x_i$ and $x_{i+1}$ (Fig.~\ref{fig:stochs}C). If the lengths
of these arcs are $l_1,l_2,\ldots,l_{n}$, then the sum 
\begin{equation}
	B =l_1^2+l_2^2+\ldots+l_n^2
	\label{B}
\end{equation} 
grows monotonically from its smallest value $B_{\min}=n(L/n)^2=L^2/n$, produced when the points $x_k$ lay
equidistantly from each other, to its largest value, $B_{\max}=L^2$, attained when all elements share the same
location, with the mean $B^{\ast}= B_{\min}2n/(n+1)\approx 2B_{\min}$ \cite{ArnoldB1,ArnoldB2,ArnoldB3,ArnoldB4}.

Intuitively, orderly arrangements appear if the elements ``repel''  each other, ``clumping'' is a sign of 
attraction, while independent elements are placed randomly. Hence the ratio $\beta = B/B_{\min}$ can be used
to capture the orderliness of patterns:
\begin{equation}
	\begin{cases}
		\,\,\beta(X)\approx1, \,\,\, & \mbox{indicates atypically ordered, nearly equidistant sequences;} \\ 
		\,\,\beta(X)\approx\beta^{\ast}\approx2 \,\,\, & \mbox{marks statistically typical, commonly scattered sequences;}\\
		\,\,\beta(X)\gg\beta^{\ast} \,\,\, & \mbox{corresponds to clustering sequences.}
	\end{cases}
	\label{betas}
\end{equation}

\textit{6. The length $L$ of the circle} accommodating a random sample sequence $X$ in Arnold's method was 
selected so that the distance between the end points, $x_0$ and $x_n$, became equal to the mean arc length
between the remaining pairs of neighboring points, $$l_n=|x_n-x_0|_{\mod L}=\frac{1}{n-1}\sum_{i=1}^{n-1}l_i.$$


\begin{figure}[h]
	\centering
	\includegraphics[scale=0.75]{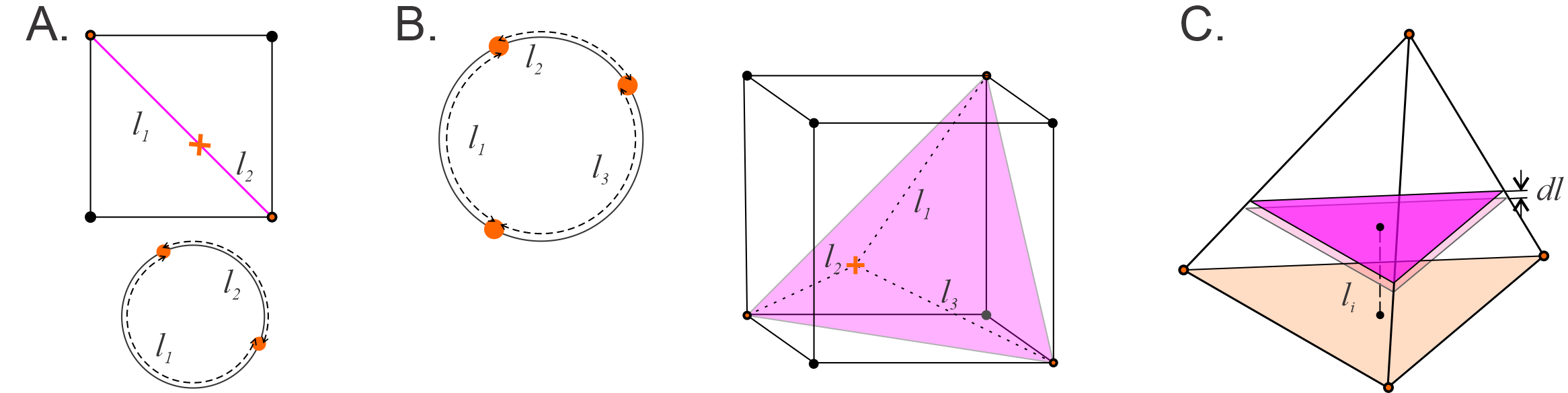}
	\captionsetup{width=.9\linewidth}
	\caption{{\footnotesize
			\textbf{Averaging over a simplex}. \textbf{A}. If two coordinates $l_1$ and $l_2$ of a two-element
			sequence could independently vary between $0$ and $L$, then the pair $(l_1,l_2)$ would cover a 
			$2D$ square. However, if the elements $(x_1,x_2)$ remain on a circle (orange dots below) then the 
			equation (\ref{L}) restricts $(l_1,l_2)$-values to the cube's diagonal (orange cross on the top panel),
			i.e., to a $1$-dimensional simplex.
			\textbf{B}. A configuration of three points on a circle corresponds to a point on the diagonal section
			of a $L$-cube.
			\textbf{C}. Tetrahedron---a section of a $4D$ cube---is the highest dimensional ($3D$) depictable 
			simplex $\sigma^{(3)}$, which is used to schematically represent $n$-dimensional simplexes, $\sigma^{(n)}$. 
			Averaging over $l_i^2$ in (\ref{Lmean}) involves integrating over it the $(n-1)$-dimensional layers of
			$\sigma^{(n)}$.
	}}
	\label{fig:cub}
\end{figure}

	\textit{7. Mean Arnold stochasticity score}. A short derivation of $\beta^{\ast}$ is provided below for 
	completeness, following the exposition given in \cite{ArnoldB4}.

	\begin{itemize}[leftmargin=0.34cm]
		\item The $n$ arcs lengths $l_1,l_2,\ldots,l_n$ produced by $n$ points, $X=\{x_1,x_2,\ldots,x_n\}$, can
		be viewed as the ``coordinates" of $X$ in a $n$-dimensional sequence space.	If these coordinates could
		vary independently on a	circle of length $L$, then the sequences would be in one-to-one correspondence
		with the points of a $n$-dimensional hypercube with the side $L$. However, since the sum of $l_i$s must
		remain fixed,
		\begin{eqnarray}
			l_1+l_2+...+l_n=L,
			\label{L}
		\end{eqnarray}
		the admissible $l$-values occupy a hyperplane that cuts between the vertices $(0,0,\ldots,0)$ and $(L,
		L,\ldots,L)$. For example, the two-element sequences described by the coordinates $l_1$ and $l_2=L-l_1$
		correspond to the points on the diagonal of a $L$-square (Fig.~\ref{fig:cub}A) and three-element sequences
		correspond to the points of a ``diagonal" equilateral triangle in the $L$-cube (Fig.~\ref{fig:cub}B). 
		The four-element sequences are represented by the points of a regular tetrahedron (Fig.~\ref{fig:cub}C)
		and so forth. Thus, a generic $n$-sequence is represented by a point in a polytope spanned by $n$ 
		vertices in $(n-1)$-dimensional Euclidean space---a $(n-1)$-\textit{simplex}, $\sigma^{(n-1)}$ 
		\cite{Alexandrov}.
		
		\item The \textit{defining property} of a simplex is that any sub-collection of its vertices spans a
		sub-simplex: a tetrahedron, $\sigma^{(3)}$, is spanned by four vertices, any three of which span a triangle 
		$\sigma^{(2)}$---a ``face" of $\sigma^{(3)}$; any two vertices span an edge, $\sigma^{(1)}$, between them,
		etc. \cite{Alexandrov}. Correspondingly, a generic section of the $\sigma^{(n-1)}$-simplex by a hyperplane
		is also a $\sigma^{(n-2)}$-simplex (Fig.~\ref{fig:cub}C).
		
		\item Averaging the sum (\ref{B}) requires evaluating the mean of each $l_i^2$,
		\begin{equation}
			\langle l_i^2\rangle=\frac{1}{V_{n-1}}\int_{\sigma^{(n-1)}} l_i^2 dV,
			\label{Lmean}
		\end{equation}
		for $i=1,2,\ldots,n$. Here $V_{n-1}$ refers to the volume of $\sigma^{(n-1)}$ and ``$dV$" refers to the
		volume of a thin layer positioned at a distance $l_i$ away from the $i^{\textrm{th}}$ face (Fig.~\ref{fig:cub}C).
		By the defining property of simplexes mentioned above, the base of this layer is a $(n-2)$-simplex 
		specified by the equation
		\begin{equation*}
			\sum_{j\neq i}^n l_j=L-l_i,
		\end{equation*}
		which implies that the sides of this base have length $L-l_i$ (just as the sides of $\sigma^{(n-1)}$
		defined	by (\ref{L}) have length $L$). The volume of the thin layer is $dV=C(L-l_{i})^{n-2}dl_i$, so
		that
		\begin{equation*}
		\langle l_i^2\rangle=\frac{C}{V_{n-1}}\int_{0}^{L} l_i^2(L-l_{i})^{n-2}dl_i=\frac{CL^{n+1}}{V_{n-1}}
			\int_0^1 u^2(1-u)^{n-2}du,
		\end{equation*}
		where $u=l_i/L$. Using the variable $v=1-u$ the latter integral yields:
		\begin{equation*}
			\langle l_i^2\rangle=\frac{CL^{n+1}}{V_{n-1}}\int_0^1 (1-v)^2v^{n-2}dv=C\left(\frac{1}{n-1}-
			\frac{2}{n}+\frac{1}{n+1}\right).
		\end{equation*}
		The volume of the  $\sigma^{(n-1)}$-simplex is
		\begin{equation*}
			V_{n-1}=\int_0^LdV=\frac{CL^{n-1}}{n-1};
			\nonumber
		\end{equation*}
		hence the sum (\ref{B}) divided by $B_{\min}=L^2/n$ yields
		\begin{equation}
			\beta_n=n^2\left(1-2\frac{n-1}{n}+\frac{n-1}{n+1}\right)=2\frac{n}{n+1}\approx \beta^{\ast}=2.
			\label{dbb}
		\end{equation}
	\end{itemize}
	\textit{8. Probability distributions of $\beta$-values} form a family parameterized by the number of
	elements in the sequence. As shown on Fig.~\ref{fig:pbeta}, these distributions have a well-defined peak
	at $\beta_{n}\approx 2\frac{n}{n+1}$ (see below) and rapidly decay as $\beta$ approaches $1$ or for 
	$\beta>3.5$, which illustrates that typical $\beta$-values, for all $n$, remain near the impartial
	mean $\beta^{\ast}\approx 2$.

\begin{wrapfigure}{c}{0.55\textwidth}
	\centering
	\includegraphics[scale=0.458]{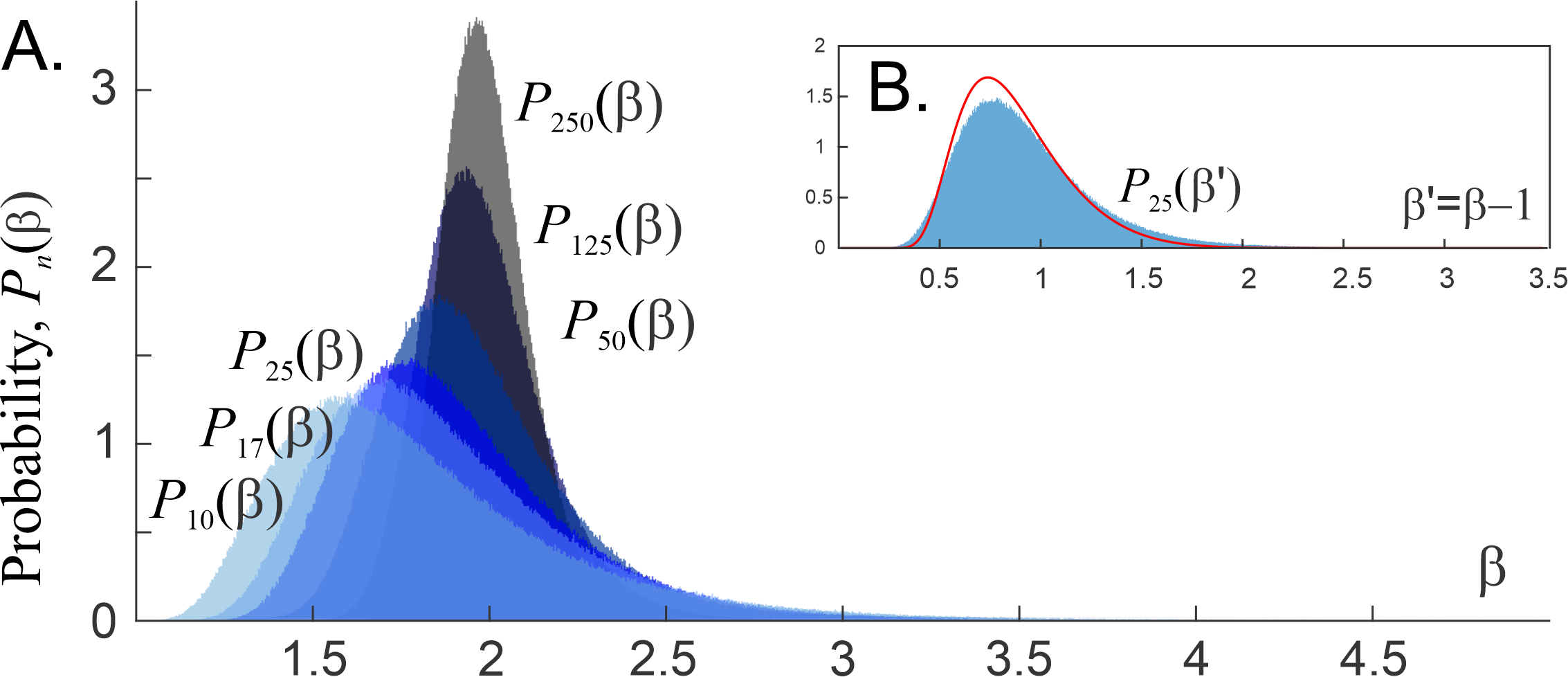}
	\caption{{\footnotesize
			\textbf{$\beta$-distributions}. \textbf{A}. Histograms of $\beta$-values obtained for $10^6$
			sequences containing $n=17$, $25$, $50$, $125$ and $250$ elements peak in a vicinity of the 
			impartial mean $\beta^{\ast}$ and rapidly decay for $\beta\lesssim 1.5$ and $\beta\gtrsim 3.5$.
			\textbf{B}. The distribution of $\beta'=\beta-1$ for sequences containing about $n=25$ points
			is close to the universal Kolmogorov distribution $P(\lambda)$ (red line, Fig.~\ref{fig:stochs}A). 
	}}
	\label{fig:pbeta}
\end{wrapfigure}

	\textit{9. The sliding window algorithm} can be implemented in two ways:
	\begin{itemize}[nosep,leftmargin=0.34cm]
		\item using a \textit{fixed window} that may capture different numbers of events at each step, i.e.,
		$L_t=L$, but $n_t$ and $n_{t+1}$ may differ;
		\item using a \textit{fixed number of events} per window, i.e., $n_t=n$, but $L_t$ and $L_{t+1}$ may differ.
	\end{itemize}
	
	In both cases, the mean window width $\bar{L}$ is proportional to the mean separation between nearest events, 
	\begin{equation}
		\bar{l}\equiv\langle l_i\rangle=\langle x_{i+1}-x_i\rangle,
		\label{meangap}
	\end{equation} 
	and the mean number $\bar{n}$ of the data points in the sample sequence, $\bar{L}=\bar{l}\bar{n}$.
	The resulting estimates for $\lambda(t)$ and $\beta(t)$ are nearly identical and, for qualitative assessments,
	 can be used interchangeably.
	 
	 \textit{10. Local averaging}. To build the dependencies between local averages, we ordered the values 
	 assumed by the independent variable, e.g., the speeds, from smallest to largest, $$\{s_1,s_2,\ldots\}\to
	 \{s'_1\leq s'_2\leq\ldots\},$$ subdivided the resulting sequence into consecutive groups containing $100$
	 elements and averaged each set, 
	  \begin{equation*}
	 	\hat{s}_i=\frac{1}{100}\sum_{k=-50}^{50}s'_{i+k}.
	 \end{equation*}
	Since each $s_{i_k}$ is associated with a particular moment of time $t_{i_k}$, we computed the averages of 
	the corresponding dependent variable, e.g., $\lambda_{i_k}$,
	 \begin{equation*}
	 	\hat{\lambda}_i=\frac{1}{100}\sum_{k=1}^{100}\lambda_{i_k}.
	 \end{equation*}
Similarly, ordering the $\beta$-scores and evaluating their local means produces the $\hat{\beta}_i$-values,
along with the means of their $\lambda_{i_k}$-counterparts that occur at the corresponding moments $t_{i_k}$,
yielding $\hat{\beta}\textrm{-}\hat{\lambda}$ dependence.


\newpage
\section{Supplementary Figures}

\renewcommand{\figurename}{Suppl. Fig.}
\setcounter{figure}{0}
\begin{figure}[h]
	\centering
	\includegraphics[scale=0.95]{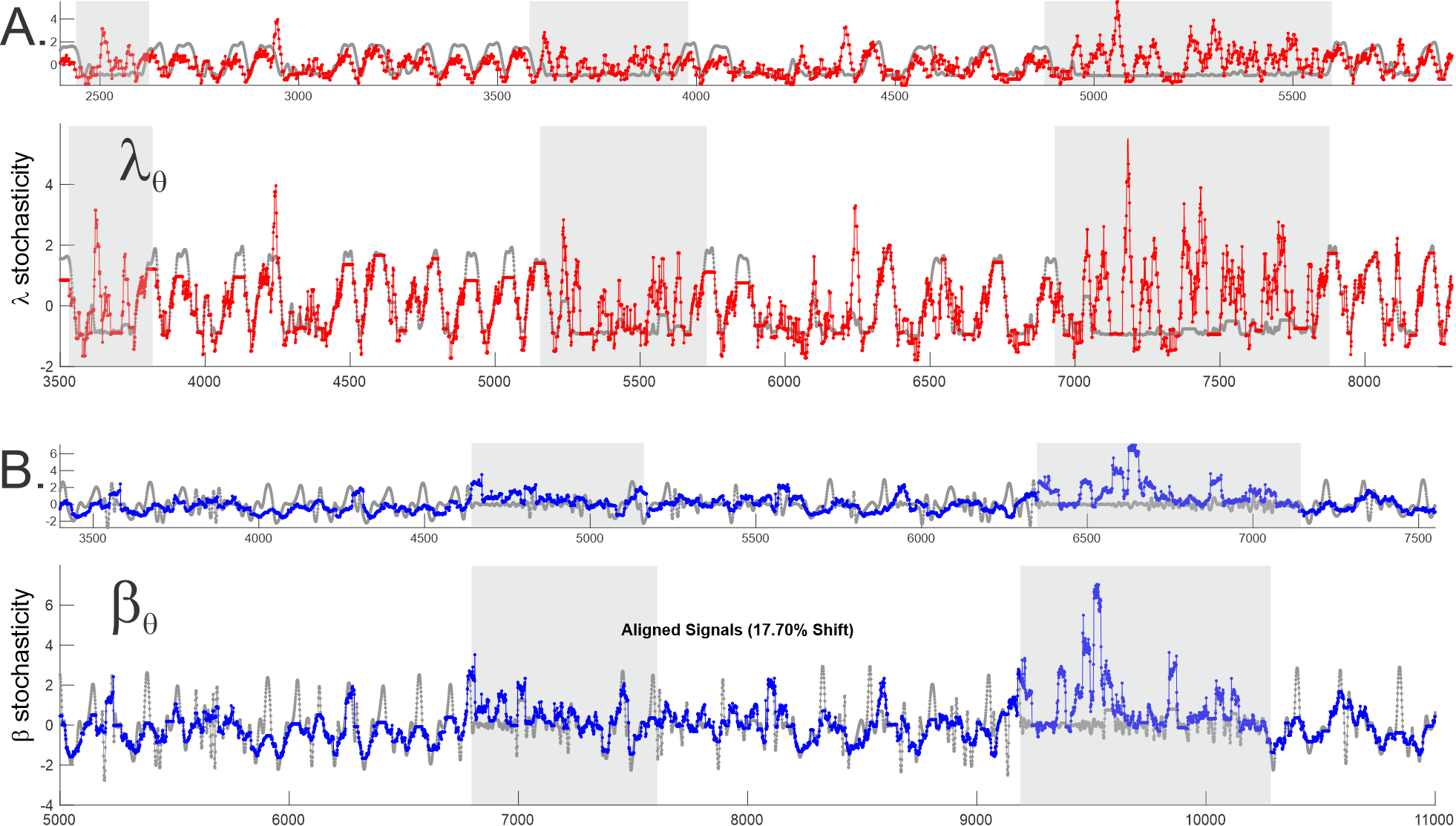}
	\caption{{\footnotesize
		\textbf{Matching waves using DTW}. \textbf{A}. Top panels show the original shapes of the speed ($s(t)$,
		gray trace) and the $theta$-wave's Kolmogorov stochasticity score ($\lambda_{\theta}(t)$, red trace).
		Bottom panel shows same functions, matched up by a sequences of local DTW-stretches. Clearly, the speed
		and the	$\lambda_{\theta}$-stochasticity have the same qualitative shape during active behavior, while
		during inactive moves (domains marked by light gray stripe) the connection is lost. 
		\textbf{B}. Same analyses carried for the mouse's acceleration ($a(t)$, gray trace) and Arnold 
		stochasticity parameter ($\beta_{theta}$, blue trace). The net amount of stretch required to match speed
		and $\lambda_{\theta}$ in this case (including the inactivity periods) is 26$\%$, while the net stretch 
		matching $\beta_{\theta}$ and the acceleration is $\sim 18\%$.
	}}
	\label{fig:bela}
\end{figure}



\renewcommand{\figurename}{Suppl. Fig.}
\begin{figure}[h]
	\centering
	\includegraphics[scale=0.8]{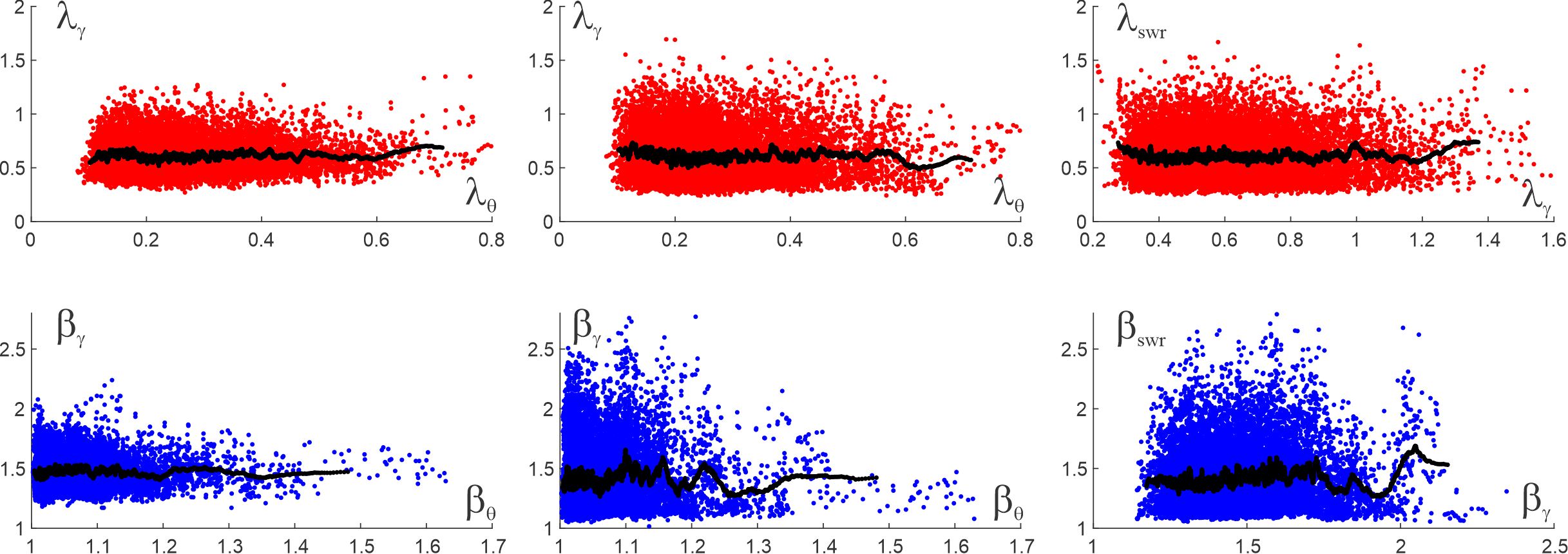}
	\caption{{\footnotesize
			\textbf{Coupling between stochasticity parameters of different waves}. 
			\textbf{A}. The geometric layout of points with coordinates $(\lambda_{\theta},\lambda_{\gamma})$, 
			$(\lambda_{\theta},\lambda_{\swr})$ and $(\lambda_{\gamma},\lambda_{\swr})$ shows that a given
			$\lambda$-value produced by one wave may pair with any $\lambda$-value that another wave is capable
			of producing, i.e., wave patterns deviate from their respective means largely independently from 
			each other. Correspondingly, the locally averaged scores $\hat{\lambda}$ lay approximately 
			horizontally, at the level of the corresponding means $\langle\lambda_{\theta}\rangle$, $\langle
			\lambda_{\gamma}\rangle$ and $\langle\lambda_{\swr}\rangle$ (see Fig.~\ref{fig:thstoch}C, 
			Fig.~\ref{fig:gstoch}B and Fig.~\ref{fig:swr}B). 
			\textbf{B}. The $\beta$-scores reveal similar lack of coupling between waves. Changes in 
			locally averaged $\hat{\beta}$-values of one wave not entrain consistent 
			$\hat{\beta}$-changes of another wave. Thus, the (dis)orderliness of one wave does not
			enforce the (dis)orderliness of the other and the stochasticity dynamics discussed above provide 
			independent characterizations of of the LFP waves.
	}}
	\label{fig:bb}
\end{figure}



\newpage
\clearpage

\section{References}

\begin{thebibliography}{99}
\bibitem{Hebb} Hebb, D. \textit{The organization of behavior: A neuropsychological theory}. J. Wiley; Chapman \& Hall (1949).
\bibitem{Syntax} Buzs\'aki, G. Neural syntax: cell assemblies, synapsembles, and readers. \textit{Neuron} \textbf{68}: 362-385 (2010).
\bibitem{Skaggs} Skaggs, W., McNaughton, B., Wilson, M. \& Barnes, C. Theta phase precession in hippocampal neuronal populations and the compression of temporal sequences. \textit{Hippocampus} \textbf{6}: 149-172 (1996).
\bibitem{Benchenane} Benchenane, K., Peyrache, A., Khamassi, M., Tierney, P., Gioanni, Y., Battaglia, F. \& Wiener, S. Coherent Theta Oscillations and Reorganization of Spike Timing in the Hippocampal- Prefrontal Network upon Learning. \textit{Neuron} \textbf{66}: 921-936 (2010).
\bibitem{Nikoli} Nikoli, D., Fries, P. \& Singer, W. Gamma oscillations: precise temporal coordination without a metronome. \textit{Trends Cogn Sci}. \textbf{17}: 54-55 (2013).
\bibitem{ClgMsr} Colgin, L., Denninger, T., Fyhn, M., Hafting, T., Bonnevie, T., Jensen, O., Moser, M-B. \& Moser, E. Frequency of gamma oscillations routes flow of information in the hippocampus. \textit{Nature} \textbf{462}: 353-357 (2009).
\bibitem{ColginGm} Colgin, L. \& Moser E. Gamma oscillations in the hippocampus. \textit{Physiology} \textbf{25}: 319-329 (2010).
\bibitem{LisBuz} Lisman, J. \& Buzs\'aki, G. A neural coding scheme formed by the combined function of gamma and theta oscillations. \textit{Schizophr. Bull}. \textbf{34}(5): 974-80 (2008). 
\bibitem{Lismn1} Lisman, J. \& Jensen, O. The Theta-Gamma Neural Code. \textit{Neuron} \textbf{77}: 1002-1016  (2013).
\bibitem{Lismn3} Lisman, J. The theta/gamma discrete phase code occurring during the hippocampal phase precession may be a more general brain coding scheme. \textit{Hippocampus} \textbf{15}: 913-922 (2005).

\bibitem{BuzTheta2} Buzs\'aki, G. Theta rhythm of navigation: link between path integration and landmark navigation, episodic and semantic memory. \textit{Hippocampus}, \textbf{15}: 827-840 (2005).
\bibitem{Richard} Richard, G., Titiz, A., Tyler, A., Holmes, G., Scott, R. \& Lenck-Santini, P. Speed modulation of hippocampal theta frequency correlates with spatial memory performance. \textit{Hippocampus}, \textbf{23}(12): 1269–1279 (2013). 
\bibitem{Kropff} Kropff, E., Carmichael, J., Moser, E. \& Moser, M.-B. Frequency of theta rhythm is controlled by acceleration, but not speed, in running rats. \textit{Neuron}, \textbf{109}: 1-11 (2021). 

\bibitem{Rangel} Rangel, L., Rueckemann, J., Riviere, P., Keefe, K., Porter, B., Heimbuch, I., Budlong, C. \& Eichenbaum, H. Rhythmic coordination of hippocampal neurons during associative memory processing. \textit{eLife} \textbf{5}: e09849 (2016).

\bibitem{Wilkinson} Wilkinson, C. \& Nelson, C. Increased aperiodic gamma power in young boys with Fragile X Syndrome is associated with better language ability. \textit{Molecular Autism}. \textbf{12}(1): 17 (2021).

\bibitem{Donoghue} Donoghue, T., Haller, M., Peterson, E., Varma, P., Sebastian, P., Gao, R., Noto, T., Lara, A., Wallis, J., Knight, R., Shestyuk, A. \& Voytek, B. Parameterizing neural power spectra into periodic and aperiodic components. \textit{Nature Neuroscience}. \textbf{23}(12): 1655-1665 (2020).
\bibitem{Eissa} Eissa, T., Tryba, A., Marcuccilli, C., Ben-Mabrouk, F., Smith, E., Lew, S., Goodman, R., McKhann, G., Frim, D., Pesce, L., Kohrman, M., Emerson, R., Schevon, C. \& van Drongelen, W. Multiscale Aspects of Generation of High-Gamma Activity during Seizures in Human Neocortex. \textit{eneuro}. \textbf{3}(2): ENEURO.0141-15.2016 (2016).
\bibitem{Nicola} Nicola, W. \& Clopath, C. A diversity of interneurons and Hebbian plasticity facilitate rapid compressible learning in the hippocampus. \textit{Nature Neuroscience}. \textbf{22}(7): 1168-1181 (2019).
\bibitem{Hawkins} Hawkins, S., Situational influences on rhythmicity in speech, music, and their interaction. \textit{Philosophical Transactions of the Royal Society B: Biological Sciences}. 369(1658): 20130398 (2014).
\bibitem{Blanco} Blanco, A. \& Ramirez, R. Evaluation of a Sound Quality Visual Feedback System for Bow Learning Technique in Violin Beginners: An EEG Study. \textit{Frontiers in Psychology}. \textbf{10}(165) (2019).
\bibitem{Cole} Cole, S. \& Voytek, B. Brain oscillations and the importance of waveform shape. \textit{Trends in Cognitive Sciences} \textbf{21}(2): 137-149 (2017).

\bibitem{Will} Will, U. \& Berg, E. Brain wave synchronization and entrainment to periodic acoustic stimuli. \textit{Neurosci Lett}. \textbf{424}(1): 55-60 (2007). 
\bibitem{Mostafa} Mostafa, H., M\"uller, L. \& Indiveri, G. Rhythmic Inhibition Allows Neural Networks to Search for Maximally Consistent States. \textit{Neural Computation}. \textbf{27}(12): 2510-2547 (2015).

\bibitem{Kolmogorov} Kolmogorov, A. Sulla determinazione empirica di una legge di distribuzione. \textit{Giornale dell'Istituto Italiano degli Attuari,} \textbf{4}(1): 83-91 (1933).
\bibitem{Stephens} Stephens, M. Introduction to Kolmogorov (1933) On the Empirical Determination of a Distribution. In: Kotz S., Johnson N.L. (eds) Breakthroughs in Statistics. Springer Series in Statistics (Perspectives in Statistics). Springer, New York, NY (1992). 
\bibitem{Arnold1} Arnold, V. Orbits' statistics in chaotic dynamical systems. \textit{Nonlinearity} \textbf{21}: T109 (2008).
\bibitem{Arnold2} Arnold, V. Empirical study of stochasticity for deterministic chaotic dynamics of geometric progressions of residues. \textit{Funct. Anal. Other Math.} \textbf2: 139-149 (2009).
\bibitem{Arnold3} Arnold, V. To what extent are arithmetic progressions of fractional parts stochastic? \textit{Russian Mathematical Surveys} \textbf{63}: 205 (2008).
\bibitem{Arnold4} Arnold, V. Stochastic and deterministic characteristics of orbits in chaotically looking dynamical systems. \textit{Trans. Moscow Math. Soc}.
\textbf{70}: 31–69 (2009)
\bibitem{Arnold5} Arnold, V. Measuring the objective degree of randomness of a finite set of points. Lecture at the school ``Contemporary Mathematics," (in Russian). Joint Institute for Nuclear Research, Dubna, Ratmino (2009).

\bibitem{KolMen} Kolmogoroff, A. On a new confirmation of Mendel's laws. \textit{Dokl. Akad. Nauk. USSR} \textbf{27}:37-41 (1940). 
\bibitem{Stark} Stark, A. \& Seneta, E. A.N. Kolmogorov's defence of Mendelism. \textit{Genetics and Molecular Biology}, \textbf{34}(2): 177-186 (2011).

\bibitem{Gurzadyan1} Gurzadyan, V., Yan, H., Vlahovic, G., Kashin, A., Killela, P., Reitman, Z., Sargsyan, S., Yegorian, G., Milledge, G. \& Vlahovic, B. Detecting somatic mutations in genomic sequences by means of Kolmogorov–Arnold analysis. \textit{R. Soc. open sci}. \textbf{2}: 150143 (2015).

\bibitem{Gurzadyan2} Gurzadyan, V. \& Kocharyan, A. Kolmogorov stochasticity parameter measuring the randomness in the cosmic microwave background. \textit{Astronomy \& Astrophysics}, \textbf{492}(2): L33 - L34 (2008).

\bibitem{Brandouy} Brandouy, O., Delahaye, J-P. \& Ma, L. Estimating the algorithmic complexity of stock markets. \textit{Algorithmic Finance}, \textbf{4}(3-4): 159-178 (2015).

\bibitem{Ford} Ford, K. From Kolmogorov's theorem on empirical distribution to number theory. In: Charpentier \'E., Lesne A., Nikolski N.K. (eds) \textit{Kolmogorov's Heritage in Mathematics}. Springer, Berlin, Heidelberg (2007). 

\bibitem{ArnoldB1} Arnold, V. Ergodic and arithmetical properties of geometrical progression's dynamics and of its orbits. \textit{Moscow Mathematical Journal}, \textbf{5}(1), 5–22 (2005).
\bibitem{ArnoldB2} Arnold, V. Topology and statistics of arithmetic and algebraic formulae, \textit{Russian Math. Surv}. \textbf{58}(4) 637–664 (2003).
\bibitem{ArnoldB3} Arnold, V. Euler Groups and Arithmetics of Geometric Progressions, Moscow, MCCME (2003).

\bibitem{ArnoldB4} Arnold, V. \textit{Lectures and Problems: A Gift to Young Mathematicians}, American Math Society, Providence (2015).

\bibitem{ColginR} Colgin, L. Rhythms of the hippocampal network. \textit{Nat. Rev. Neurosci.} \textbf{17}: 239-249 (2016).

\bibitem{Burgess} Burgess, N. \& O'Keefe, J. The theta rhythm. \textit{ Hippocampus} \textbf{15}: 825-826 (2005).
\bibitem{BuzTheta1} Buzs\'aki, G. Theta oscillations in the hippocampus. \textit{Neuron} \textbf{33}: 325-340 (2002). 

\bibitem{Bol1} Bol'shev, L. Asymptotically Pearson Transformations \textit{Theory of Probability \& Its Applications}, \textbf{8}(2): 121-146 (1963).
\bibitem{Vrbik1} Vrbik, J. Small-sample corrections to Kolmogorov–Smirnov test statistic. \textit{Pioneer Journal of Theoretical and Applied Statistics}, \textbf{15}(1–2): 15–23 (2018).
\bibitem{Vrbik2} Vrbik, J. Deriving CDF of Kolmogorov-Smirnov Test Statistic. \textit{Applied Mathematics}, \textbf{11}: 227-246 (2020).

\bibitem{Berndt} Berndt, D. \& Clifford, J. Using Dynamic Time Warping to Find Patterns in Time Series, Proceedings of the 3rd International Conference on Knowledge Discovery and Data Mining. AAAI Press: Seattle, WA. p. 359–370 (1994).
\bibitem{Salvador} Salvador, S. \& Chan, P. Toward accurate dynamic time warping in linear time and space. \textit{Intell. Data Anal}. \textbf{11}(5): 561–580 (2007).

\bibitem{Neamtu} Neamtu, R., Ahsan, R., Rundensteiner, E., S\'ark\"ozy, G., Keogh, E., Anh Dau, H., Nguyen, C. \& Lovering, C. Generalized Dynamic Time Warping: Unleashing the Warping Power Hidden in Point-Wise Distances. Proceedings of 34$\textrm{th}$ International Conference on Data Engineering ()ICDE), pp. 521-532 (2018)


\bibitem{Vakman} Vakman, D. \& Vainshtein, L. Amplitude, phase, frequency-Fundamental concepts of oscillation theory. \textit{Sov.Phys.Usp.}, \textbf{20}(12): 1002-1016 (1977).
\bibitem{Rice} Rice, S. Envelopes of narrow-band signals, \textit{Proc. IEEE}, \textbf{70}: 692-699 (1982).

\bibitem{Ahmed} Ahmed, O. \& Mehta, M. Running speed alters the frequency of hippocampal gamma oscillations. \textit{J. Neurosci}. \textbf{32}(21): 7373–7383 (2012). 
\bibitem{Montgomery} Montgomery, S. \& Buzsáki, G. Gamma oscillations dynamically couple hippocampal CA3 and CA1 regions during memory task performance. \textit{Proc. Natl. Acad. Sci.}, \textbf{104}(36): 14495–14500 (2007). 

\bibitem{Girardeau2} Girardeau, G., Zugaro, M. Hippocampal ripples and memory consolidation. \textit{Curr. Opin. Neurobiol} \textbf{21}, 452-459 (2001).
\bibitem{Singer} Singer, A., Carr, M. Karlsson, M. \& Frank, L. Hippocampal SWR Activity Predicts Correct Decisions during the Initial Learning of an Alternation Task. \textit{Neuron} \textbf{77}: 1163-1173 (2013).
\bibitem{Roux} Roux, L., Hu, B., Eichler, R., Stark, E. \& Buzs\'aki, G. Sharp wave ripples during learning stabilize the hippocampal spatial map. \textit{Nat. Neurosci.} \textbf{20}: 845-853 (2017). 

\bibitem{Sadowski2} Sadowski, J., Jones, M. \& Mellor, J. Sharp-Wave Ripples Orchestrate the Induction of Synaptic Plasticity during Reactivation of Place Cell Firing Patterns in the Hippocampus. \textit{Cell Reps}. \textbf{14}, 1916-1929 (2016).

\bibitem{Denovellis} Denovellis, E., Gillespie, A., Coulter, M., Sosa, M., Chung, J., Eden, U. \& Frank, L. Hippocampal replay of experience at real-world speeds. \textit{Elife} \textbf{10}:e64505 (2021). 

\bibitem{BarneSpars} Barnes, C. A., McNaughton, B. L., Mizumori, S. J., Leonard, B. W. \& Lin, L. H. Comparison of spatial and temporal characteristics of neuronal activity in sequential stages of hippocampal processing. \textit{Prog. Brain Res}. \textbf{83}: 287–300 (1990).

\bibitem{Wu} Wu, X. \& Foster, D. Hippocampal Replay Captures the Unique Topological Structure of a Novel Environment. \textit{J. Neurosci}. \textbf{34}:6459--6469 (2014).

\bibitem{Kudrimoti} Kudrimoti, H., Barnes, C., \& McNaughton B. Reactivation of hippocampal cell assemblies: 
effects of behavioral state, experience, and EEG dynamics. \textit{J Neurosci}. \textbf{19}: 4090–101 (1999).
\bibitem{ONeil} O'Neil, J., Senior, A., Allen, K., Huxter, J. \& Csicsvari, J. Reactivation of experience-dependent
cell assembly patterns in the hippocampus. \textit{Nat. Neurosci}. \textbf{11}: 209–216 (2008).
\bibitem{MosRev} Moser, E. Kropff, E. \& Moser, M.-B. Place Cells, Grid Cells, and the Brain's Spatial Representation System. \textit{Annu Rev Neurosci}. \textbf{31}(1): 69-89 (2008).
\bibitem{Nitz1} Nitz D. Tracking route progression in the posterior parietal cortex. \textit{Neuron} \textbf{49}(5):747-56 (2006). 

\bibitem{Ede} van Ede, F., Quinn, A., Woolrich, M., \& Nobre, A. Neural Oscillations: Sustained Rhythms or Transient Burst-Events? \textit{Trends in neurosciences}, \textbf{41}(7): 415–417 (2018). 
\bibitem{Jones} Jones S. When brain rhythms aren't `rhythmic': implication for their mechanisms and meaning. \textit{Curr. Opin. Neurobiol}., \textbf{40}: 72–80  (2016). 

\bibitem{Fransen} Fransen, A. van Ede, F. \& Maris E. Identifying neuronal oscillations using rhythmicity, \textit{NeuroImage} \textbf{118}: 256-267 (2015).

\bibitem{Neumann} Neumann, J., Wigner, E. \& Hofstadter, R. \textit{Mathematical foundations of quantum mechanics}. Princeton, Princeton University Press (1955).

\bibitem{Mises} Mises, B. v., Grundlagen der Wahrscheinlichkeitsrechnung. \textit{Mathematische Zeitschrift} \textbf{5}: 52-99 (1919).
\bibitem{Uspenskii} Uspenskii, V., Semenov A. \& Shen', A. Can an individual sequence of zeros and ones be random? \textit{Russ Math. Surveys}, \textbf{45}(1): 105-162 (1990).
\bibitem{Volchan} Volchan, S. What Is a Random Sequence? The American Mathematical Monthly. \textbf{109}(1): 46-63 (2002).

\bibitem{Berger1} H. Berger, \"Uber das Elektrenkephalogramm des Menschen. \textit{Archiv f\"ur Psychiatrie und Nervenkrankheiten} \textbf{98}: 231–254; \textbf{99}: 555–574 (1933). 
\bibitem{Christoph} Christoph, A. On some questions of V.I. Arnold on the stochasticity of geometric and arithmetic progressions. \textit{Nonlinearity} \textbf{28}: 3663 (2015).

\bibitem{Weiss1} Weiss, M. Non-Gaussian Properties of the EEG During Sleep. \textit{Electroencephalography and Clinical Neurophysiology}, \textbf{34}: 200-202 (1973).
\bibitem{Weiss2} Weiss, M. Testing EEG data for statistical normality. Images of the Twenty-First Century. \textit{Proceedings of the Annual International Engineering in Medicine and Biology Society}. \textbf{2}: 704-705 (1989). 
\bibitem{Weiss3} Weiss, M. Testing correlated ``EEG-like" data for normality using a modified Kolmogorov-Smirnov statistic. \textit{IEEE Trans Biomed Eng}. \textbf{33}(12):1114-20. (1986). 
\bibitem{McEwen} McEwen, J. \& Anderson, G. Modeling the Stationarity and Gaussianity of Spontaneous Electroencephalographic Activity. \textit{IEEE Transactions on Biomedical Engineering}, \textbf{22}: 361-369 (1975).
\bibitem{Greenstein} Greenstein, Y., Pavlides, C. \& Winson, J. Long-term potentiation in the dentate gyrus is preferentially induced at theta rhythm periodicity. \textit{Brain Research}. \textbf{438}(1): 331-334 (1988).
\bibitem{Hinder} Hinder, M., Goss, E., Fujiyama, H., Canty, A., Garry, M., Rodger, J. \& Summers, J. Inter- and Intra-individual Variability Following Intermittent Theta Burst Stimulation: Implications for Rehabilitation and Recovery. \textit{Brain Stimulation}. \textbf{7}(3): 365-371 (2014).

\bibitem{Larson} Larson, J. \& Munk\'acsy, E. Theta-burst LTP. \textit{Brain Research} \textbf{1621}: 38-50 (2015).
\bibitem{Sheridan} Sheridan, G., Moeendarbary, E., Pickering, M., O'Connor, J. \& Murphy, K. Theta-Burst Stimulation of Hippocampal Slices Induces Network-Level Calcium Oscillations and Activates Analogous Gene Transcription to Spatial Learning. \textit{PLOS ONE} \textbf{9}: e100546 (2014).
\bibitem{Nguyen} Nguyen, P. \& Kandel, E. Brief theta-burst stimulation induces a transcription-dependent late phase of LTP requiring cAMP in area CA1 of the mouse hippocampus. \textit{Learning \& Memory} \textbf{4}: 230-243 (1997).

\bibitem{Lundqvist} Lundqvist, M., Rose, J., Herman, P., Brincat, S. L., Buschman, T. \& Miller, E. Gamma and Beta Bursts Underlie Working Memory. \textit{Neuron}, \textbf{90}(1): 152–164 (2016). 

\bibitem{Replays} Babichev, A., Morozov, D. \& Dabaghian, Y. Replays of spatial memories suppress topological fluctuations in cognitive map. \textit{Network Neuroscience} \textbf{3}(2): 1-18 (2019).

\bibitem{ChengJi} Cheng, J. \& Ji, D. Rigid firing sequences undermine spatial memory codes in a neurodegenerative mouse model. \textit{eLife}, \textbf{2}: e00647 (2013). 
\bibitem{Ciupek} Ciupek, S., Cheng, J., Ali, Y., Lu, H., \& Ji, D. Progressive functional impairments of hippocampal neurons in a tauopathy mouse model. \textit{J. Neurosci}. \textbf{35}(21): 8118–8131 (2015). 

\bibitem{Alexandrov} Aleksandrov, P. \textit{Elementary concepts of topology}. F. Ungar Publishing (1965).

\end{thebibliography}
\end{document}